\documentclass[lettersize,journal]{IEEEtran}
\usepackage{amsmath,amsfonts}
\usepackage{algorithmic}
\usepackage{algorithm}
\usepackage{array}
\usepackage[caption=false,font=normalsize,labelfont=sf,textfont=sf]{subfig}
\usepackage{textcomp}
\usepackage{stfloats}
\usepackage{url}
\usepackage{verbatim}
\usepackage{graphicx}
\usepackage{cite}

\usepackage{booktabs}
\usepackage{tabularx}
\usepackage{xcolor,colortbl}
\usepackage{multirow}
\usepackage{import}
\usepackage{soul}

\usepackage{hyperref}
\usepackage[capitalise]{cleveref}
\definecolor{highlight}{RGB}{230, 245, 230}
\usepackage{array}      
\usepackage{enumitem}   
\usepackage{ragged2e}   

\begin{document}

\title{The Vertical Challenge of Low-Altitude Economy: Why We Need a Unified Height System?}

\author{Shuaichen YAN, Xiao HU, Jiayang SUN, Zeyuan YANG, Shipeng LI,~\IEEEmembership{Fellow,~IEEE}, Heung-Yeung SHUM,~~\IEEEmembership{Fellow,~IEEE,} Shijun YIN, Yuqing TANG
\thanks{This paper was produced by the IEEE Publication Technology Group. They are in Piscataway, NJ.}
\thanks{Manuscript received April 19, 2021; revised August 16, 2021.}

\thanks{Shuaichen YAN is with the Department of Urban Planning and Design, The University of Hong Kong, 999077, China.}

\thanks{Xiao Hu, Jiayang SUN, Zeyuan YANG, Shipeng LI, Yuqing TANG are with the Lower Airspace Economy Research Institute, International Digital Economy Academy, ShenZhen, 5100085, China(Corresponding author: Yuqing TANG, e-mail: tangyuqing@idea.edu.cn).}

\thanks{Heung-Yeung SHUM is with the Hong Kong University of Science and Technology, Hong Kong, 999077, China.}

\thanks{Shijun YIN is with the Chinese Society of Aeronautics and Astronautics, Beijing, 100012, China.}

}

\markboth{Journal of \LaTeX\ Class Files,~Vol.~14, No.~8, August~2021}%
{Shell \MakeLowercase{\textit{et al.}}: A Sample Article Using IEEEtran.cls for IEEE Journals}


\maketitle

\begin{abstract}

The explosive growth of the low-altitude economy, driven by eVTOLs and UAVs, demands a unified digital infrastructure to ensure safety and scalability. However, the current aviation vertical references are dangerously fragmented: manned aviation relies on barometric pressure, cartography uses Mean Sea Level (MSL), and obstacle avoidance depends on Above Ground Level (AGL). This fragmentation creates significant ambiguity for autonomous systems and hinders cross-stakeholder interoperability. In this article, we propose Height Above Ellipsoid (HAE) as the standardized vertical reference for lower airspace. Unlike legacy systems prone to environmental drift and inconsistent datums, HAE provides a globally consistent, GNSS-native, and mathematically stable reference. We present a pragmatic bidirectional transformation framework to bridge HAE with legacy systems and demonstrate its efficacy through (1) real-world implementation in Shenzhen's partitioned airspace management, and (2) a probabilistic risk assessment driven by empirical flight logs from the PX4 ecosystem. Results show that transitioning to HAE reduces the required vertical separation minimum, effectively increasing dynamic airspace capacity while maintaining a target safety level. This work offers a roadmap for transitioning from analog height keeping to a digital-native vertical standard.
\end{abstract}

\begin{IEEEkeywords}
Height System,
  Low Altitude Economy,
  Advanced Air Mobility,
  Unmanned Aircraft System Traffic Management, 
  Height Above Ellipsoid,
  Lower Airspace Management.
\end{IEEEkeywords}

\section{Introduction}
\IEEEPARstart{L}{ower} airspace, the airspace below $1,000$ meters Above Ground Level (AGL), is rapidly emerging as a critical enabler of urban air mobility (UAM), logistics, and emergency services. Projections estimate the global low-altitude economy (LAE) will surpass $1.5 \ trillion$ by 2030, driven by innovations in electric Vertical Take-off and Landing (eVTOL) aircraft and unmanned aerial systems (UAS)~\cite{idea-1}. However, the absence of a unified vertical reference system jeopardizes this growth. Recent incidents, such as near-misses between UAVs and helicopters in Shenzhen and regulatory conflicts in the EU’s U-space framework~\cite{sesar_conops_2024}, underscore the risks posed by inconsistent height referencing. Although latitude and longitude benefit from universal standardization, height systems remain fragmented, relying on disparate references such as barometric pressure, geoid models, or terrain data. This fragmentation complicates airspace integration, collision avoidance, and regulatory compliance, particularly in densely populated urban areas.

Existing studies have focused on airspace classification~\cite{bauranov2021designing}, traffic management~\cite{garrow2022proposed,pons2022understanding,hamissi2023survey,elsayed2024impact,liao2021views, liao2023views,quan2020low}, reduced vertical separation minima~\cite{schumann2017static}, and policy frameworks~\cite{caac2023}, yet few address the foundational challenge of height standardization. For instance, Bauranov and Rakas~\cite{bauranov2021designing} highlight the role of digital twins in UAM but overlook vertical reference inconsistencies, while Hamissi and Dhraief~\cite{hamissi2023survey} detail UAS traffic management without resolving height ambiguities. Traditional systems like Mean Sea Level (MSL) and AGL face limitations: MSL relies on regional datums prone to geoid deviations~\cite{pavlis2012development}, while AGL depends on terrain models that are often outdated or restricted~\cite{pavlis2012development}. Barometric Q-codes, although extensively used in manned aviation, suffer from atmospheric variability~\cite{berberan1997barometric}, thus rendering them unreliable for autonomous systems. 
These gaps hinder the safe integration of UAVs, eVTOLs, and legacy aircraft in shared airspace. 
The European H2020 ICARUS project~\cite{icarus2022,icarus2020,icarus2024} represents the most comprehensive effort to address vertical reference fragmentation in low-level airspace. It proposes a U-space service architecture that integrates GNSS-derived geometric altitude with legacy barometric systems. However, ICARUS prioritizes pressure-based references to ensure interoperability with general aviation. This emphasis results in limited exploration of non-barometric systems, which are expected to play a significant role in lower airspace operations.

Our contributions advance the field by:
\begin{itemize}
    \item \textbf{Critical Analysis:} Exposing unresolved gaps in height system interoperability and stakeholder coordination that threaten the scalability of Urban Air Mobility (UAM).
    \item \textbf{Methodological Innovation:} Introducing a practical framework for adopting HAE as the standard height system, including forward/backward transformation models to ensure backward compatibility with legacy aviation.
    \item \textbf{Empirical Validation:} Quantifying the safety and economic impact of HAE using rigorous \textit{Reich Collision Risk} and \textit{Erlang-B} capacity models, parameterized by real-world sensor error distributions derived from PX4 flight logs.
\end{itemize}
By resolving vertical reference ambiguities, this work supports scalable, safe, and economically viable lower airspace operations, advancing global LAE initiatives.
The remainder of this paper is structured as follows: ~\cref{sec:preliminaries} reviews commonly used height systems, including their definitions, reference surfaces, and measurement methodologies. ~\cref{sec: status} analyzes the current challenges of height system usage in lower airspace, emphasizing operational ambiguities and stakeholder coordination issues.~\cref{sec: hae} proposes HAE as the standardized vertical reference system, demonstrating its advantages in computability, stability, and interoperability.\cref{sec: airspace} validates HAE's efficacy through two case studies: partitioned airspace management in Shenzhen and a quantitative safety/capacity analysis based on empirical error modeling. Finally, ~\cref{sec:con} concludes the paper, summarizes key findings, and outlines future research directions for refining HAE--based frameworks.

\section{Preliminaries}
\label{sec:preliminaries}
Height is theoretically defined as the metric distance from a point to a reference surface, measured along a clearly defined geometric or physical path such as the perpendicular line to a plane (e.g., geometric height) or the plumb line aligned with gravity (e.g., orthometric height relative to the geoid). While this definition aligns with the intuitive notion of vertical distance, it introduces ambiguity due to two factors: the choice of measurement path (geometric vs. physical) and the reference surface itself, which may vary between global geodetic datums (e.g., ellipsoids or geoids), local tangent planes, or regional vertical datums.
These variations result in context-dependent interpretations of height, thereby requiring precise terminological clarification.
Though often used interchangeably, ``height'' is the general term, while ``elevation'' usually refers to the vertical distance of a ground point above mean sea level, and “altitude” denotes an aircraft’s height above mean sea level (e.g. geoid-referenced orthometric heights), typically used in operational and navigational contexts.
By establishing these distinctions, this section aims to ensure consistency in subsequent discussions of height systems, emphasizing the importance of standardized definitions for accurate interpretation across applications.

\subsection{Height Above Ellipsoid}
Height above ellipsoid, or simply HAE, also referred to as geodetic height, is the vertical distance measured relative to a reference ellipsoid--a mathematically defined surface approximating Earth’s shape. Unlike localized vertical datums, HAE offers global consistency because its reference ellipsoid is standardized worldwide, ensuring uniformity in measurements across regions and applications. This compatibility with mainstream geodetic coordinate systems (e.g., WGS84, ITRF) simplifies integration with modern positioning technologies, making HAE a practical choice for global navigation, aerial surveying, and digital mapping.

HAE values are directly obtainable from GNSS such as GPS, BeiDou, GLONASS, and Galileo. Advanced techniques like Real-Time Kinematic (RTK) and Precise Point Positioning (PPP) enhance measurement accuracy to centimeter-level precision. While each GNSS may adopt a slightly different reference ellipsoid or frame (e.g., GPS uses WGS84, Galileo employs GTRF), these systems are rigorously defined, enabling precise conversions between height measurements using well-established formulas with quantifiable errors. For instance, transformation parameters between the WGS84 and ITRF ensure mutual consistency, with discrepancies typically below 10 cm--a tolerance sufficient for most aviation and geospatial applications~\cite{Malys2018}. Applying time-dependent transformation models, such as geocentric position vector adjustments, can further minimize residual deviations.

The primary limitation of HAE is its inability to directly represent height relative to terrain or sea level, necessitating conversions for human-centric applications.

\subsection{Height Above Mean Sea Level and Orthometric Height}
Height Above Mean Sea Level, or AMSL/MSL in short, describes the vertical distance from a point to a regionally averaged sea level, which serves as its reference datum. In practice, MSL is established through decades of tide gauge observations to account for tidal fluctuations and seasonal variations. For instance, China’s 1985 National Height Datum relies on tide gauge data collected at Qingdao from 1952 to 1979~\cite{QDHY198803001}, while the United States uses the North American Vertical Datum of 1988 (NAVD 88), anchored to local sea level conditions~\cite{noaa_navad88}. A key advantage of MSL lies in its physical intuitiveness: it directly reflects sea level as observed in nature, making it indispensable for coastal engineering and floodplain mapping, where altitude above terrain is critical. Furthermore, MSL-based systems are deeply entrenched in national surveying traditions, ensuring compatibility with historical datasets. However, MSL has significant limitations: it varies regionally due to ocean currents, tectonic uplift, and climate-driven sea level rise, necessitating country-specific datums. Additionally, MSL can deviate from the global geoid—a gravity-derived equipotential surface approximating mean sea level—by up to several meters~\cite{noaa_ngs_glossary}, complicating global standardization.

Another height system commonly considered alongside MSL is orthometric height--the vertical distance from a point to the geoid, measured along the gravity-aligned plumb line~\cite{meyer2006does}--offers a globally consistent framework. Unlike MSL, orthometric height integrates both geometric and physical properties, as it aligns with Earth’s gravity field. Modern tools like the Earth Gravitational Model 2008 (EGM2008)~\cite{pavlis2012development}, which synthesizes satellite gravimetry (e.g., GRACE and GOCE missions) with terrestrial data, enable precise geoid modeling at centimeter-level accuracy. This capability supports applications such as earthquake deformation monitoring and precision agriculture, as demonstrated by China’s recent adoption of a 3D elevation system combining GNSS and geoid models~\cite{most_china_2024}. The primary advantage of orthometric height is its global interoperability: it transcends regional MSL discrepancies, enabling seamless height conversions across borders. One drawback of orthometric systems is their reliance on the accuracy of geoid models. These models need continuous updates to accommodate gravitational changes. Additionally, obtaining orthometric heights requires access to precise geoid models, but such access may be restricted.

While MSL and orthometric height systems differ in their reference surfaces, they are increasingly interconnected. Regional MSL datums are often aligned to the geoid through geopotential adjustments, bridging the gap between local tradition and global standardization. 
Looking ahead, advancements in GNSS technology and next-generation gravitational models (e.g., EGM2020) are shifting the focus toward orthometric systems, which offer inherent compatibility with satellite positioning.

\subsection{Height Above Ground Level}
Height above ground level or AGL in short refers to the vertical distance from a point to the natural terrain surface directly beneath it, reflecting true topographic elevation.
By definition, AGL employs a Digital Terrain Model (DTM) as its reference surface, explicitly excluding artificial structures (e.g., buildings) and vegetation to represent bare ground. 
However, in contexts where surface features are relevant—such as urban UAV navigation--a Digital Surface Model (DSM), which includes buildings and trees, may instead define the ground surface. This variability in reference surfaces creates inherent inconsistency: AGL measurements depend entirely on local terrain or surface models, meaning the ``ground level'' is spatially unique for every horizontal coordinate.

AGL’s critical role in airspace management is evident in global regulations. For instance, NASA’s Unmanned Aircraft System Traffic Management (UTM) restricts UAVs to 400 feet AGL~\cite{faa_traffic_management}, the European Union’s U-space framework caps operations at 150 meters AGL (or above the highest structure)~\cite{sesar_conops_2024}, and China’s Interim Regulations on UAV Flight Management impose a 120-meter AGL ceiling for micro, light, and small UAVs operating within approved airspace.~\cite{gov_cn_policy_2023}. 
These rules highlight AGL’s importance in ensuring safety and minimizing collisions in low-altitude airspace. Despite its operational utility, AGL poses significant practical challenges. 
\begin{itemize}
    \item First, achieving measurement accuracy is notoriously challenging. While radio altimeters, LiDAR, or ultrasonic sensors can directly measure AGL, their reliability degrades in complex environments. In urban areas, multipath interference—where signals reflect off buildings rather than reaching the ground—skews readings. 
    \item Second, AGL relies on high-fidelity DTM data to standardize measurements across locations. However, DTMs face two major hurdles: (1) many governments restrict DTM publication due to national security concerns, limiting accessibility, and (2) rapid urbanization or natural terrain shifts can render DTMs outdated within months. 
    \item Third, the lack of a unified reference surface complicates height comparisons: two AGL values from different locations cannot be directly compared without knowing their respective terrain models.
\end{itemize}
These challenges propagate into operational risks. For example, inconsistent AGL data impedes regulatory compliance checks, height conflict detection, and airspace deconfliction systems. A UAV operating in a city with outdated DSM data might incorrectly assume obstacle clearance, while another in a restricted zone could inadvertently breach airspace limits due to inaccurate DTM-based calculations. 

Further complicating these issues, although AGL is critical for terrain-referenced safety protocols, its dependency on localized, mutable reference surfaces and fragile measurement workflows limits its practicality as a height system for airspace management, especially considering the dynamic nature of lower airspace--with increasing UAV traffic and evolving infrastructure.

\subsection{Q-codes: QFE, QNH, and QNE}
Altitude and atmospheric pressure share an inverse predictable relationship: pressure $P(h)$ decreases exponentially with increasing altitude $h$, as described by the barometric formula derived from hydrostatic equilibrium and the ideal gas law~\cite{berberan1997barometric}. This relationship underpins barometric altimeters, which estimate altitude by measuring ambient pressure. In civil aviation, these instruments have been indispensable for decades due to their simplicity and reliability. However, pressure-based altitude systems face a critical limitation: atmospheric pressure varies spatially and temporally due to weather patterns, temperature fluctuations, and geographic location. For instance, a low-pressure weather system or a temperature inversion can skew altimeter readings by hundreds of feet, making it challenging to compare measurements across regions or times.

To mitigate inconsistencies, the International Civil Aviation Organization (ICAO) introduced the Q-code system, which standardizes pressure references for altimeters\cite{prentice2010aviation}:
\begin{itemize}
\item QFE denotes the atmospheric pressure at the elevation of the aerodrome or runway threshold, which can be utilized to indicate the aircraft's height relative to that point as the reference system on the ground with calibrating their barometric altimeters. During takeoff and landing, QFE height is important.
\item QNH is an altimeter sub-scale setting to obtain elevation when on the ground, which adjusts the local barometric pressure to MSL under standard International Standard Atmosphere (ISA) conditions as the reference system. This adjustment allows pilots to determine their altitude above sea level, which is essential for ensuring safe clearance over terrain, especially during approach and departure phases. QNH height also corresponds with the elevation data found on aeronautical charts.
    \item QNE fixes the altimeter to the standard pressure of 1013.25 hPa (ISA sea-level pressure), providing a uniform reference for cruising altitudes. This prevents midair collisions by ensuring all aircraft at the same flight level use identical baseline pressure.
\end{itemize}
Despite the Q-code system’s utility, pressure-based altitude determination remains inherently unstable. Localized weather changes require constant QNH/QFE updates, and discrepancies persist between regions due to non-standard atmospheric conditions. For example, a pilot transitioning from a high-pressure to a low-pressure zone without updating QNH risks flying lower than indicated. These shortcomings are particularly acute in lower airspace. Consequently, while QFE and QNH remain in ICAO’s documentation\cite{icao_codes_2014}, they are being phased out in favor of GNSS-based systems, which provide globally consistent, weather-independent altitude data.

\begin{figure*}
    \centering
    \includegraphics[width=1\textwidth]{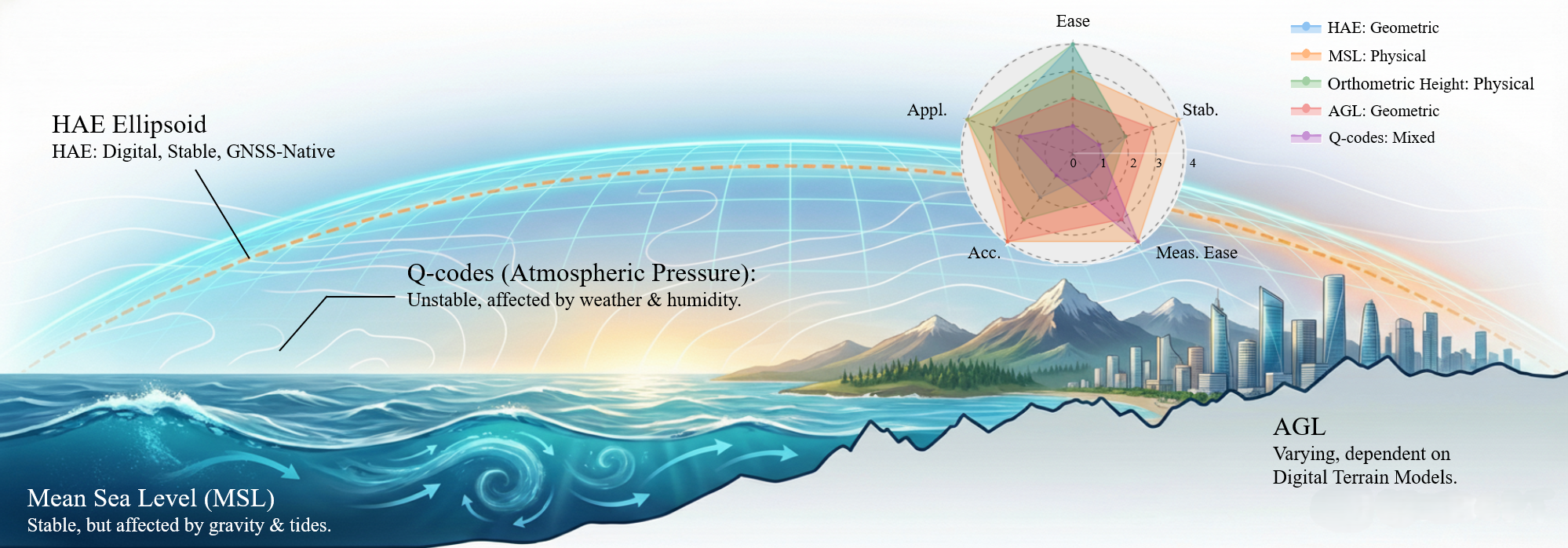}
    \caption{Comparative Evaluation of Height Systems: \textbf{Scale:} 1 (lowest) to 4 (highest). Abbreviations: \textit{Ease} for Conceptual simplicity, \textit{Stab.} for Reference stability, \textit{Meas. Ease} for Measurement practicality, \textit{Acc.} for Measurement accuracy, and \textit{Appl.} for Practical applications.}
    \label{tab:height_system_comparison}
\end{figure*}

\subsection{Comparison of Height Systems}
Height systems are evaluated across six criteria: ease of conceptual understanding, reference surface stability, measurement practicality, accuracy, geometric/physical basis, and practical applications, scored on a 1–4 scale.
Ease of understanding reflects how intuitive the concept is for users or professionals. 
Stability refers to the long-term consistency of the reference surface and its resistance to external influences. 
Measurement ease (Meas. Ease) reflects how easily the height system can be measured in practice. This includes the simplicity or complexity of required instruments and measurement techniques. 
Measurement accuracy (Acc.) assesses the precision of the system under typical conditions. It focuses on the ability to directly measure height without relying on indirect methods or conversions. 
Geometric versus physical height (Geom./Phys.) indicates whether the system represents purely geometric positioning, relates to the Earth's physical properties, or combines both. Instead of numerical scores, this aspect is labeled as "Geom.," "Phys.," or "Mixed." Finally, the extent of practical applications (Appl.) measures how frequently the system is used in various fields and real-world scenarios. 
As shown in Table~\ref{tab:height_system_comparison}, HAE excels in stability and direct GNSS measurability but requires abstract geometric interpretation. In contrast, MSL is highly intuitive yet vulnerable to climate-driven sea-level shifts. Orthometric heights achieve peak accuracy via gravity-aligned geoid models but demand technical expertise, while AGL prioritizes simplicity but suffers from unstable terrain references. Q-codes, though easy to measure, are weather-sensitive. 
\section{The Fragmentation Problem: Why Current Systems Fail UAM?}
\label{sec: status}

\begin{figure}
    \centering
    \includegraphics[width=1\linewidth]{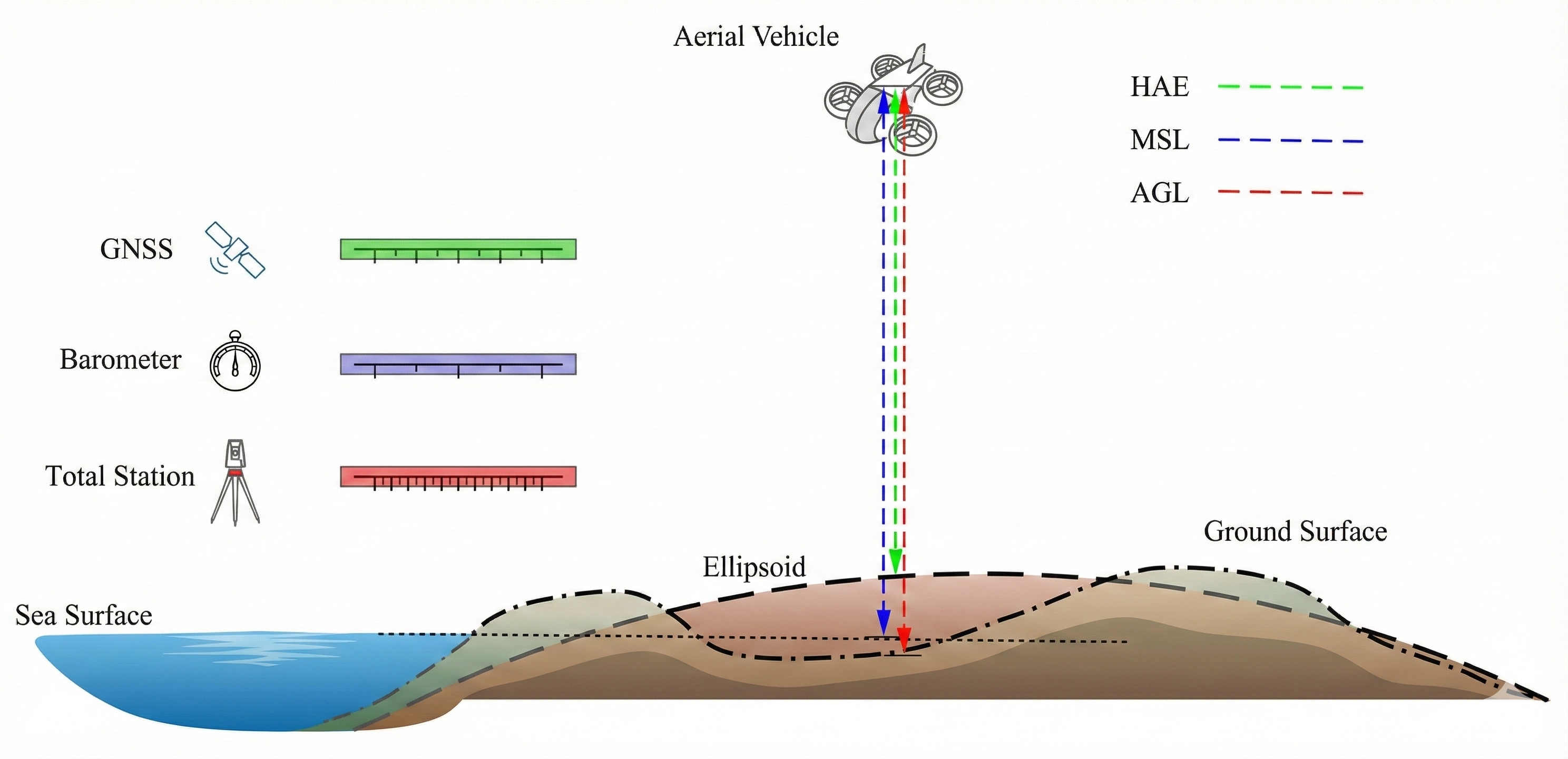}
    \caption{Illustration of de-facto height systems for LAE.}
    \label{fig:enter-height}
\end{figure}
Lower airspace is a valuable and multifaceted resource, characterized by its diverse range of activities and stakeholders. Effective airspace development and management require a collaborative approach involving multiple stakeholders, including airspace authorities, users, and service providers. This section provides an overview of the current status of height usage in lower airspace, highlighting the usually overlooked challenges posed by the lack of a unified height system.

\subsection{Multiple Stakeholders in Lower Airspace}
\label{sec:Properties_lae}
Lower airspace is a complex domain involving various stakeholders, each with distinct roles and requirements. These stakeholders include:
\begin{itemize}
    \item \textbf{Authorities}: The authorities involve multiple departments and institutions, such as those related to civil aviation, natural resources, social economy and urban planning. Examples include the Federal Aviation Administration (FAA) in the United States and the Civil Aviation Administration of China (CAAC). These authorities are tasked with managing airspace, overseeing flight operations, facilitating data exchange, and establishing regulatory frameworks essential for the effective governance of airspace. However, there is an inconsistency in the height systems used within the authorities. While regulation documents related to lower airspace typically use AGL, the fundamental geographic data is mostly referenced in MSL or AGL, which poses challenges for inner data integration.
    \item \textbf{Airspace Users}: This category includes private companies, government entities, and individual users who conduct diverse flight activities such as aerial photography, cargo delivery, and urban air mobility. The use of different height systems by these users, such as Q-codes for helicopters and HAE for UAVs and eVTOLs, poses coordination challenges.
    \item \textbf{Service Providers}: These entities support airspace authorities and users by providing essential services. Examples include U-Space Service Providers, Common Information Service Providers, Supplemental Data Service Providers, CNS Service Providers, Air Traffic Service Providers, and Aeronautical Information Service Providers (AISP)~\cite{sesar_conops_2024}. These service providers receive, operate, and disseminate information involving heights, making it crucial to ensure consistency in height systems.
\end{itemize}

\subsection{Challenges of Inconsistent Height Description and Missing Conversion Framework}
The diverse stakeholders outlined in Section 3.1, combined with the absence of a standardized height system in lower airspace, give rise to two key issues affecting low-altitude activities and operators.
\begin{itemize}
    \item \textbf{Inconsistent Height Descriptions}:  
    The lack of precise and consistent height descriptions results in misunderstandings and misinterpretations. Unlike latitude and longitude, height is not always explicitly linked to its vertical reference or type in the everyday language. The differences between height systems, such as vertical datum, measurement methods, errors, and units, can be significant. For example, the MSL used in geodesy and the ``altitude'' shown by a barometric altimeter are not the same. Misunderstandings can arise when technical distinctions are overlooked. This issue is further compounded by the fact that different stakeholders use different height systems, leading to confusion and potential safety risks.
    For example, two major UAV management systems in China--UTMISS and UOM--use different height systems. UTMISS uses the 1985 National Vertical Datum System, while UOM uses ``true height'', a concept similar to AGL. 
    \item \textbf{Inconsistent Height Data and Lack of Relevant Conversion Scheme}: 
    The lack of a standardized height system directly results in the misuse of various height measurements in low-altitude activities. While some height systems have rigorous definitions or models, others rely on reference surfaces or parameters that are not easily accessible or are drifting across space and time. For example, DTM data used for AGL measurements and Q-codes heights based on atmospheric pressure are often inaccessible or variable. Without a standardized height system and a proper conversion scheme, these data are not comparable to height data from other systems, hindering data integration and flight management. 
\end{itemize}
These issues have significant implications across multiple dimensions of lower airspace management.
\begin{itemize}
    \item \textbf{Flight Safety}: 
Lower airspace is often crowded with buildings, power lines, and natural features like trees and hills. Precise height control is crucial for navigating these obstacles safely. Without accurate and consistent height measurement, the risk of collision increases, potentially leading to property damage or personal injury. Height measurement is also crucial for the safety of operations involving multiple UAVs or between UAVs and manned aircraft, particularly in congested urban areas or near airports.

    \item \textbf{Operation Efficiency}: 
    The existing chaotic height systems in lower airspace will restrict the fine use of the airspace, which in the end degrades the the efficiency of UAV operations. A standardized height system enables the division of airspace into finer and more precise vertical layers, increasing its capacity and enabling multiple UAVs or aircraft to operate safely within the same general area but at different heights. This optimization of airspace use translates into faster deliveries, more flights per day, and higher economic values in industries like logistics.

     \item \textbf{Height Systems Coordination}: 
    Sufficient height description and comparability are essential for coordinating paths and trajectories before and during flights. For example, UTM systems from different sectors are being developed to manage flight operations in lower airspace. These systems must exchange height information across various stakeholders, including regulators, air traffic controllers, UAV operators, and manned aircraft pilots. Standardized height measurement ensures a common language for height measures, promoting smoother integration and information sharing across different systems.

    \item \textbf{Regulatory Compliance}: 
    Authorities are developing regulations for the safe integration of UAVs into national airspace systems. These regulations often include specific guidelines for airspace classification, necessitating a unified approach to height measurement. Non-compliance with standardized height regulations can lead to fines, operational delays, or even bans on UAV operations in certain regions. Standardized height measurement simplifies regulatory compliance, ensuring safer and more reliable UAV operations.
\end{itemize}

As LAE continues to expand rapidly, the adoption of standardized height measurement systems will become increasingly essential. This work is motivated by the aim to take a small yet meaningful step toward establishing a practical proposal that lays the foundation for safer, more efficient, and better-integrated lower airspace management in the future.

\section{HAE: The Digital-Native Vertical Reference}
\label{sec: hae}
In response to the challenges of low-altitude activities listed above, it is evident that a standardized height system is essential. After thoroughly analyzing the various height systems, we propose that the HAE system is the most suitable for meeting the diverse requirements of different stakeholders. HAE offers a balance of precision, consistency, and ease of integration across platforms, addressing the needs of authorities such as aviation and surveying entities. In the following sections, we present a detailed analysis of why HAE is the optimal choice and how it effectively addresses the challenges posed by other height systems.

\subsection{Criteria for an Ideal Lower Airspace Height System}
Defining a height system for lower airspace requires balancing theoretical rigor with operational practicality. A similar question in the field of geodesy has been extensively studied for decades, leading to a modernization movement that ultimately adopted HAE as a de-facto standard~\cite{meyer2006does}. 
A series of works by~\cite{meyer2006does} systematically investigates different height systems commonly used in the field of geodesy~\cite{meyer2006does}. However, their research lacks consideration of height systems relevant to UAVs and aviation, making it less applicable to lower airspace. This gap in understanding height systems across different domains is a key driving force behind our work, as we aim to address the unique challenges posed by lower airspace and provide a height system that meets the needs of all stakeholders involved in this emerging field. 
As the height system would be used for airspace management and data exchange, we propose three criteria essential for a modern lower airspace height system:
\begin{itemize}
    \item \textbf{Digital Compatibility}: 
    Heights must be computationally tractable to enable automated airspace management. As emphasized in~\cite{idea-2}'s work~\cite{idea-2}, only a \textit{digital-native} management system can handle the high-density, high-frequency, and high-complexity activities associated with lower airspace. 
    \item \textbf{Reference Stability}: 
    The vertical datum must remain temporally and spatially consistent to ensure long-term usability. A stable reference should be resist to environmental drift, enable backward compatibility with legacy systems, and simplify regulatory compliance by decoupling measurements from transient parameters.
    \item \textbf{Operational Accuracy and Accessibility}: 
    The measurement protocols must balance precision with practicality. While high precision is theoretically achievable with any height system, it is contingent upon the simplicity of the measurement process. To ensure practical applicability, the associated methods must avoid excessively complex or costly procedures while still meeting acceptable accuracy standards. Additionally, the devices used for such measurements must be cost-effective and widely available, facilitating broad adoption across different sectors. By striking a balance between precision and simplicity, it becomes possible to reliably gather height data in a cost-effective manner, thereby optimizing the efficiency and functionality of the height system.
\end{itemize}
These criteria address the limitations of legacy systems while aligning with the low-altitude economy’s need for interoperability, safety, and scalability. By prioritizing digital compatibility, stability, and pragmatic accuracy, we lay the groundwork for a height system that serves both UAVs and traditional aviation.

\subsection{HAE: the Optimal Choice}
Based on the aforementioned properties, we now show why HAE outperforms other height systems:
\subsubsection{Digital Compatibility}
HAE represents inherent digital compatibility and interoperability in height systems. As a purely geometric system tied to a globally standardized ellipsoid, HAE enables seamless integration with GNSS positioning, algorithmic data processing, and real-time conversions to other vertical datums (e.g., MSL via geoid models or AGL via DTMs). This computability is critical for airspace management systems that must reconcile data across stakeholders, such as UAV operators using HAE with authorities relying on MSL-based charts.

In contrast, barometric heights suffer from spatiotemporal nonlinearity. A 10 hPa pressure change could correspond to altitude shifts ranging from 30 to 100 meters, depending on weather conditions and other factors. Without real-time atmospheric models, these heights resist algorithmic integration. While MSL offers directly observed height values, the inconsistency of vertical datums and associated geoid models across various datasets necessitates the alignment of these references prior to any calculations, potentially hindering computational efficiency. Furthermore, this conversion process requires HAE as an intermediary, highlighting the superior computability of HAE. For AGL, height measurements are inherently terrain-relative, meaning that comparing AGL values requires auxiliary information, such as terrain references from DTMs or DEMs. Typically, HAE is used to transform values across different references. Additionally, because AGL depends directly on the variable terrain, using it exclusively for distance calculations over large areas is impractical.

\subsubsection{Reference Stability}
HAE is based on a reference ellipsoid, like WGS84, which remains constant unless explicitly redefined. The reference ellipsoid remains static unless deliberately updated, providing a stable reference system that is unaffected by terrain, sea level changes, or atmospheric conditions, making it particularly valuable for applications requiring long-term consistency in height measurement.

In contrast, AGL relies on dynamic terrain surfaces subject to erosion, sediment deposition, and human activity. Seasonal changes in a river delta, for instance, can shift AGL values by meters, complicating obstacle databases and flight path planning. 

\begin{figure}[!htbp]
\centering
\includegraphics[width=1\linewidth]{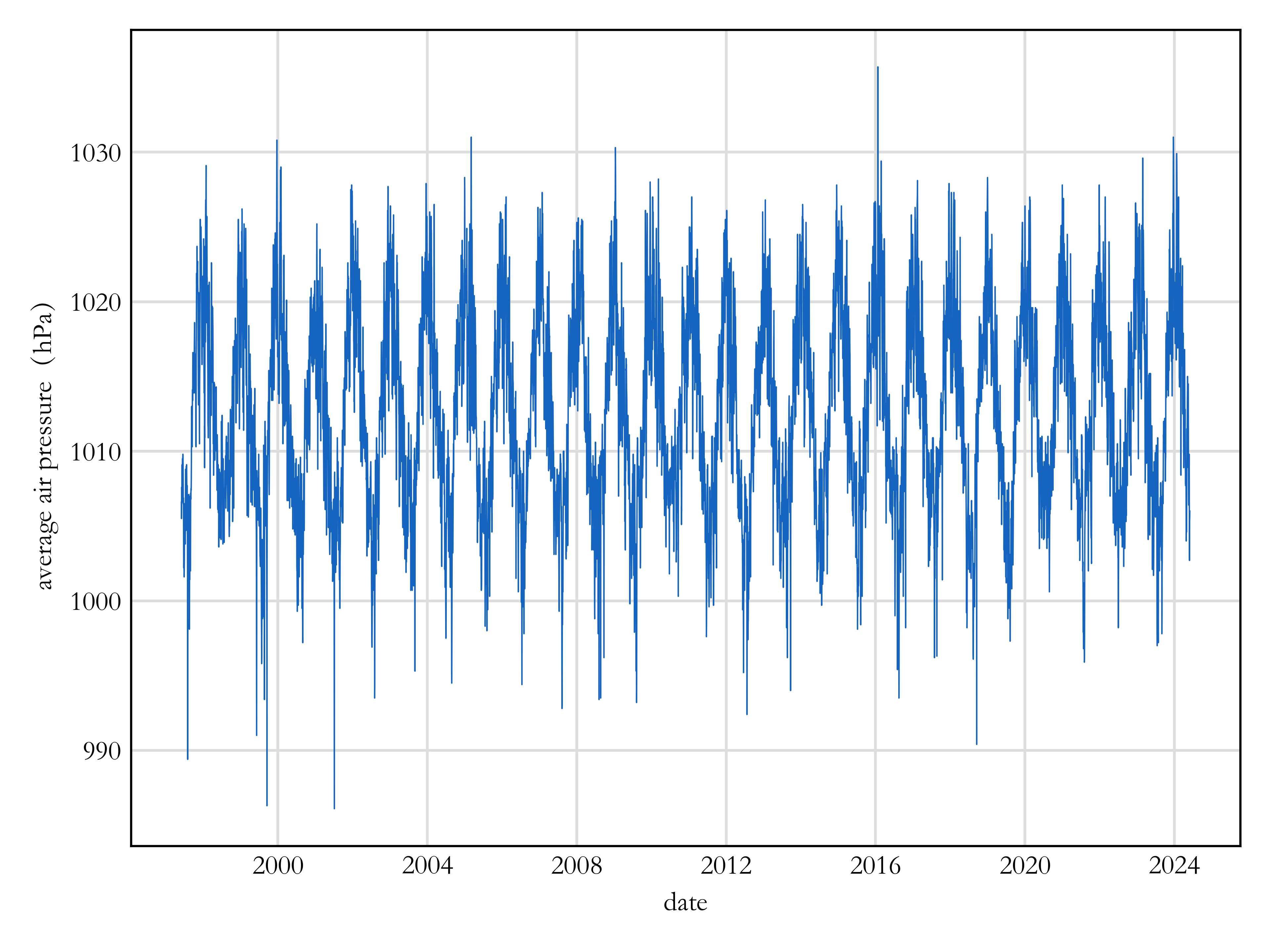}
\caption{Atmospheric pressure obtained at Hong Kong International Airport from 1997 to 2024.}
\label{fig1: pressure}
\end{figure}

Barometric heights are even less reliable as they depend on atmospheric pressure, which fluctuates spatially and temporally due to weather conditions, temperature, humidity, and other factors. To illustrate the limitations of pressure-based height measurements, we analyze the atmospheric pressure data from Hong Kong International Airport\cite{hkobs_mslp}, as presented in~\cref{fig1: pressure} and summarized in~\cref{tab:height_summary}. The dataset, spanning from June 1997 to May 2024, reveals significant daily variations in atmospheric pressure due to fluctuating meteorological conditions. These variations highlight the instability of pressure-based height references, as the baseline pressure at the airport changes frequently. Consequently, the height measurements derived from atmospheric pressure can vary significantly based on the chosen reference pressure, causing stability issues.

\begin{table}[!htbp]
    \centering
    \small
    \caption{Statistical Summary of Daily Average Atmospheric Pressure at Hong Kong International Airport (hPa). All values are in meters and SD represents Standard Deviation.}
    \label{tab:height_summary}
    \begin{tabular}{|c|c|c|c|c|c|c|}
    \hline Mean & SD & Min & Median & Max & Range \\
     \hline 1012.7 & 6.5 & 986.1 & 1012.6 & 1035.7 & 49.6 \\
     \hline
    \end{tabular}
\end{table}

\begin{table*}[!htbp]
\centering
\small
\caption{Summary of DEM Datasets.}
\begin{tabular}{|p{2cm}|p{2.5cm}|p{3cm}|p{2cm}|p{2.5cm}|p{2.5cm}|}
\hline
\textbf{Dataset Name} & \textbf{Publication Year} & \textbf{Measurement Method} & \textbf{Dataset Type} & \textbf{Geodetic Reference} & \textbf{Spatial Resolution} \\
\hline
NASA SRTM & 2007 & InSAR & DSM & EGM96 & 30 m \\
\hline
AW3D30 & 2016 & Photogrammetry & DSM & EGM96 & 30 m \\
\hline
MERIT & 2017 & 
Fusion of VFP-DEM, SRTM and AW3D Datasets
& 
Technically close to DTM
& EGM96 & 90 m \\
\hline
COPERNICUS GLO-30 & 2019 & InSAR & DSM & EGM2008 & 30 m \\
\hline
NASA DEM & 2020 & Multi-dataset Fusion & DSM & EGM96 & 30 m \\
\hline
\end{tabular}
\end{table*}

\begin{table}[!htbp]
\centering
\small 
\caption{Statistics of the difference in the DEM dataset (values in meters).}
\begin{tabular}{|p{2cm}|c|c|c|c|c|}
\hline
\textbf{Dataset} & \textbf{Mean} & \textbf{SD} & \textbf{Q1} & \textbf{Median} & \textbf{Q3} \\
\hline
NASA SRTM & 5.66 & 4.79 & 2 & 4 & 8 \\
\hline
AW3D30 & 4.94 & 4.38 & 2 & 4 & 7 \\
\hline
MERIT & 4.77 & 4.44 & 1 & 3 & 7 \\
\hline
COPERNICUS GLO-30 & 5.22 & 4.48 & 2 & 4 & 7 \\
\hline
NASA DEM & 5.79 & 4.94 & 2 & 4 & 8 \\
\hline
\end{tabular}
\end{table}

MSL, while widely adopted in traditional geodesy and aviation, lacks global uniformity, largely because different countries define their own vertical datums based on locally observed sea level data. This introduces regional discrepancies, as each country’s MSL can vary depending on the location and the period of observation. For instance, China’s 1985 vertical datum was based on measurements taken in Qingdao, which used the sea level of the Yellow Sea as its reference. Given China’s extensive coastline, using this datum as a baseline for altitude measurements in locations such as Zhejiang Province leads to notable discrepancies\cite{HYCH199504007}. Similarly, the transition from the 1956 Yellow Sea vertical datum to the 1985 Qingdao vertical datum demonstrates the challenges associated with maintaining consistency in MSL references over time. Moreover, although MSL is typically calculated as an average over a defined period, it remains subject to long-term changes driven by oceanographic conditions, tectonic shifts, and climate change~\cite{meyer2006does}. Defining MSL using a geoid model offers a more consistent global reference by accounting for Earth's gravitational field; however, even this approach is not immune to long-term variability, as the geoid itself can evolve due to shifts in Earth’s mass distribution. Updates to the geoid model, such as those seen with EGM96 and subsequent models, account for these variations in Earth's gravitational field, reflecting shifts in the geoid over time\cite{pavlis2012development}.


\begin{figure}[!htbp]
\centering
\includegraphics[width=1\linewidth]{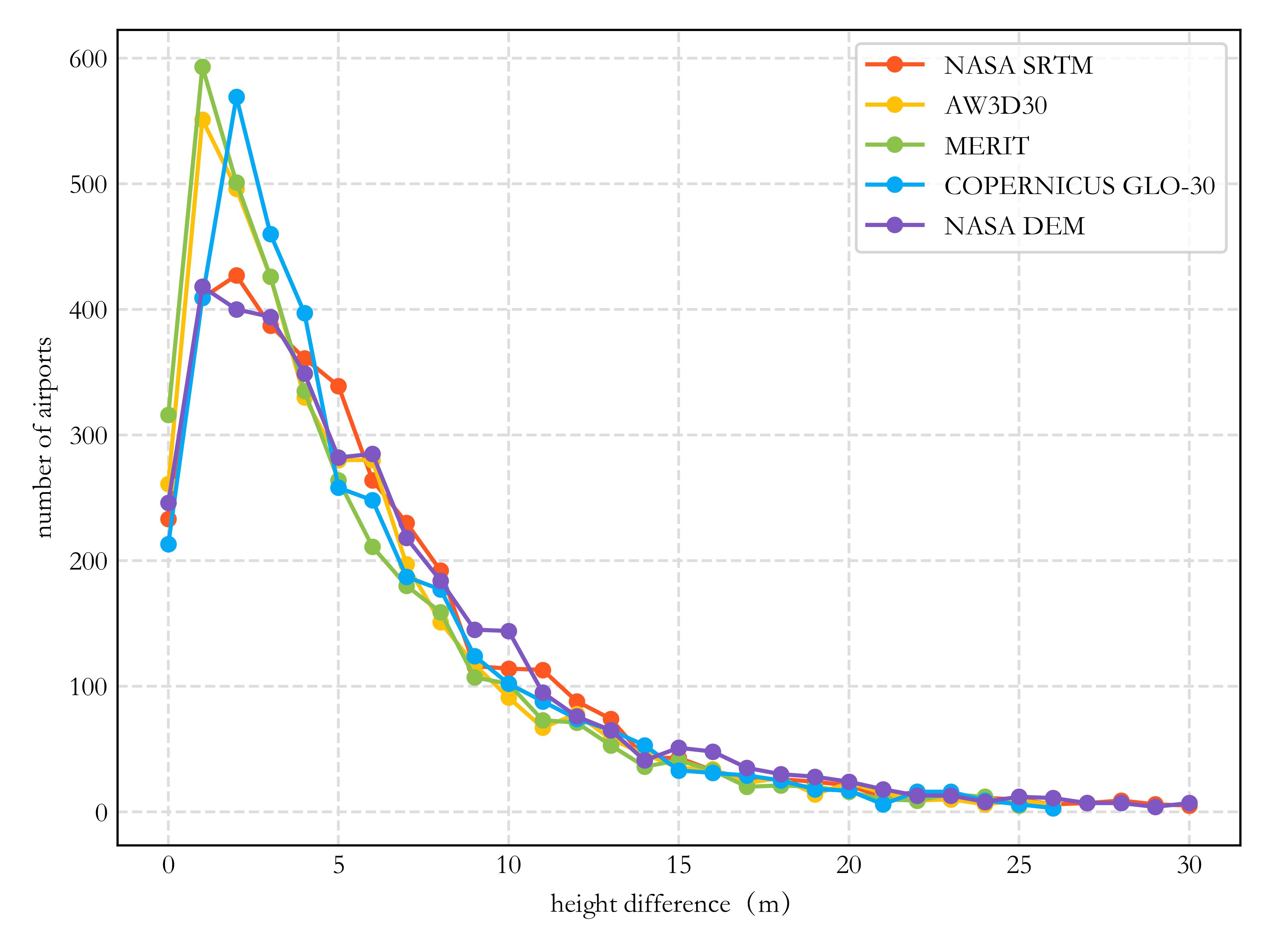}
\caption{Height differences between various DEM datasets.}
\label{fig1: dem}
\end{figure}

\subsubsection{Operational Accuracy and Accessibility}
The proliferation of GNSS technology has resolved historical accessibility barriers to HAE, enabling direct measurement with centimeter-level vertical accuracy through multi-frequency Real-Time Kinematic (RTK), Precise Point Positioning (PPP), or Processing Kinematic (PPK) techniques~\cite{schaefer2021accuracy}. Modern UAVs, eVTOLs, and helicopters now ubiquitously integrate GNSS receivers, making HAE the most accessible and precise height system for lower airspace.

In contrast, barometric systems depend on error-prone atmospheric assumptions (e.g., uniform pressure gradients, standard lapse rates) and require ancillary humidity/temperature data to approximate heights. In urban environments, these calculations are subject to significant uncertainty due to the assumptions underlying the formula, such as uniform atmospheric conditions and a standard lapse rate\cite{berberan1997barometric}. This introduces challenges for obtaining accurate height measurements.

For AGL, direct measurement methods often involve relatively expensive equipment, such as ultrasonic altimeters or radio altimeters, which are costlier compared to GNSS receivers. Although these altimeters claim high measurement accuracy, their underlying principles make it difficult to measure height above the true ground surface accurately, potentially leading to discrepancies with regulatory requirements.
Using DTM or DSM for measuring AGL or MSL relies heavily on the accuracy and recency of the DTM/DSM dataset. High-precision mapping products are often confidential due to data security reasons, limiting access to accurate AGL or MSL information. 
Open-source datasets may provide some information but often suffer from lower precision. An analysis of airport altitudes from the Global Airport Database (GADB) using various DEM datasets\cite{Partow_Airport_Database} including MERIT\cite{yamazaki2017high}, NASA DEM\cite{NASADEM_HGT}, NASA SRTM\cite{Farr2007}, COPERNICUS GLO-30\cite{Copernicus_DEM_Handbook}, and AW3D30\cite{Tadono2014} is shown in~\cref{fig1: dem} and the result demonstrate that while altitude differences are generally within 10 meters, with many within 5 meters, discrepancies still persist. These discrepancies arise from varying geoids, measurement methods, and publication times, complicating unified management for low-altitude flight operations.



In conclusion, HAE emerges as the unequivocal choice for the height system for lower airspace, offering unparalleled computability, stability, accessibility,  and accuracy compared to legacy systems like AGL, MSL, or Q-codes. The stability and clarity of its definition prevent misunderstandings and eliminate ambiguity in vertical separation, enhancing safety, efficiency, coordination capability, and simplifying regulatory compilance.

\subsection{Practical Solutions for Height System Compatibility}
A crucial consideration in proposing HAE as the standard for lower airspace involves ensuring compatibility with established height systems such as AGL, MSL, and barometric Q-codes.
Legacy systems remain deeply entrenched within aviation infrastructure, regulatory frameworks, and stakeholder workflows.
To address this, we propose a bidirectional transformation framework that ensures seamless integration of HAE with existing systems while maintaining backward compatibility.
\begin{itemize}
    \item \textbf{Forward Transformation to HAE}: To ensure compatibility, a transformation approach must be provided so that other height systems (such as AGL, MSL, or barometric height) can be seamlessly converted to HAE. This will allow existing stakeholders, who are already using other systems, to transition smoothly to HAE. The key is to provide clear and reliable transformation models for common height systems.
    \item \textbf{Backward Transformation from HAE}: For legacy systems that remain important to certain stakeholders, we must establish backward transformations so that HAE heights can be converted back to other height systems when necessary. Some stakeholders, such as air traffic controllers or geodetic authorities, may still prefer or require heights in other formats due to historical practices or specific regulations.
\end{itemize}
While these procedures sound intuitive and straightforward, they are challenging to implement in practice, especially if a height system lacks an explicit mathematical model. For example, transforming from AGL to HAE can be difficult in urban areas where DSM data is constantly changing. Similarly, converting from barometric height (Q-codes) to HAE involves complexities due to the variability of atmospheric pressure. To mitigate these challenges, we propose practical solutions for transforming between height systems in lower airspace.
\subsubsection{Conversion Models}
    For height systems where an explicit mathematical model exists, it is generally feasible to obtain a transformation that converts one height system to another. Mathematically, this can be expressed as: 
    $$
    h_{\text{out}} = f(h_{\text{in}},\ \mathcal{M})\ ,
    $$
    where $\mathcal{M}$ represents the transformation model, $h_{\text{in}}$ and $h_{\text{out}}$ denote the input and output height measurements, respectively. For instance, transformations between HAE and AGL can be achieved using geoid models such as EGM96 or EGM2008. These models approximate the height difference between AGL and HAE by providing the necessary geoid separation, with modern GNSS chips like U-Blox and Novatel supporting such transformations. The accuracy of the transformation depends on the precision of the geoid model and the quality of the input GNSS data.
    The following list a few model-based transformation examples for height systems commonly seen in lower airspace:
    
\begin{itemize}
    \item MSL $\leftrightarrow$ HAE: Based on a specific geoid or Earth's gravity field model (such as EGM2008 or EGM96), extract the corresponding geoid height \( N \) for the aircraft’s latitude and longitude. Conversion can be made forward and backward using the relationship \( h = H + N \), where \( H \) is the MSL height and $h$ is the HAE height.
    \item AGL $\leftrightarrow$ HAE: From  a DTM or 3D map, query a ground height using the aircraft’s latitude and longitude. If the query height is already a HAE measure, then the HAE height of the aircraft is directly obtained. Whereas if the ground height is in MSL format, Then apply the MSL $\leftrightarrow$ HAE routine to complete the conversion.
\end{itemize}

A concrete example is given to showcase the routine. Here’s a calculation example for the HAE value of an aircraft at Hong Kong International Airport. 
\\
\textbf{Assumptions:}
\begin{itemize}
    \item[*] Airport baseline pressure: 1005.4 hPa.
    \item[*] Aircraft pressure: 998 hPa.
    \item[*] DEM and EGM: ASTGTM DEM dataset corresponding to the EGM96 model.
    \item[*] The average temperature between two pressure levels: the airport's average temperature of 24.3°C is used.
\end{itemize}
\textbf{Steps to calculate the HAE value:}
\begin{enumerate}
    \item Calculate the HAE of the Baseline Point: Obtain the baseline point's MSL $h_\text{MSL}$ using the latitude and longitude $(\phi, \lambda)$ and the ASTGTM DEM grid. Then, get the geoid height $N$ from the EGM96 gravity field model for the airport point. Finally, compute the baseline geodetic height using the formula: $h_b=h_\text{MSL}+N$. We will get $h_\text{MSL}=4$m, $N=-3.1$m, $h_b=0.9$m.
    \item Calculate the pressure thickness between the aircraft and the baseline point, given the pressures at the airport and the aircraft, as well as the average temperature between those two pressure levels: $h_\text{QFE}=64.84$m.
    \item Calculate the aircraft’s HAE: $h=h_b+h_\text{QFE}$, so we get $h=65.74$m.
\end{enumerate}

\begin{figure}[!htbp]
\centering
\includegraphics[width=1\linewidth]{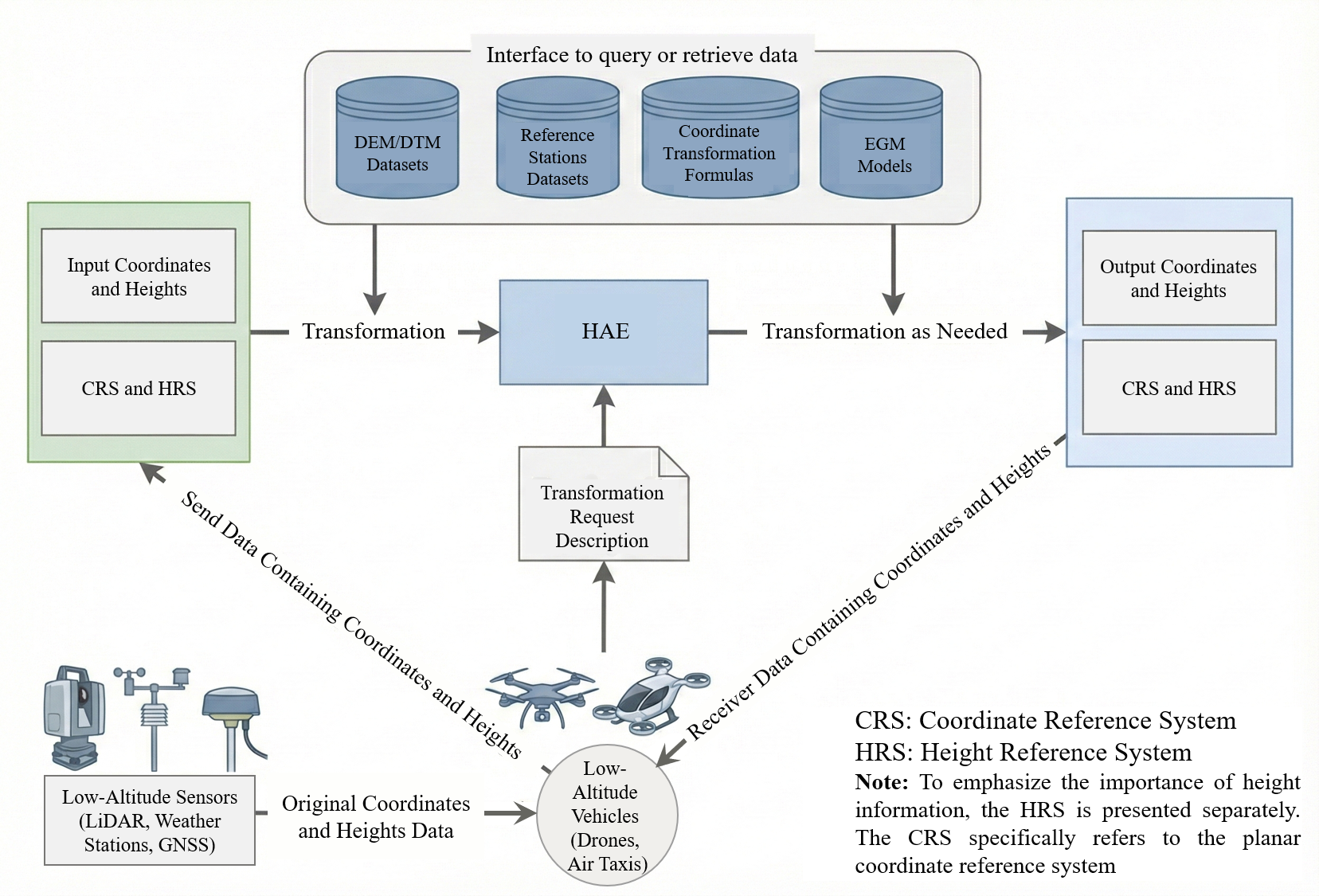}
\caption{An overview of height system transformation framework.}
\label{fig: silas_core}
\end{figure}

    \subsubsection{Calibration Methods}
    In cases where an exact transformation is impractical or undefined, calibration techniques can be employed to derive pairwise calibration values for converting one height system to another. This can be written as:
    $$
    h_{\text{out}} = f(h_{\text{in}},\ \mathcal{C})\ ,
    $$
    where $\mathcal{C}$ represents the calibration parameters, and $h_{\text{in}}$ and $h_{\text{out}}$ are the input and output height values, respectively. Calibration can be achieved by setting up a suite of instruments to simultaneously measure heights in both systems, allowing for straightforward algebraic operations to perform the conversion. For example, assuming the QNH at a calibration point (typically a runway) and the reference geodetic height HAE, estimate the aircraft’s HAE is pretty simple, i.e. Aircraft HAE = Aircraft QNH - Calibration Point QNH + Calibration Point HAE. 
    In some cases, it may be infeasible to set up such calibration systems on-site. Thus, empirical calibration parameters--derived from experience or existing data--can also be used. However, this will result in lower accuracy. For example, in aviation, barometric altitude (various Q-codes) does not provide exact height measurements, but an empirical model assumes that a 1 hPa pressure difference corresponds to approximately 8.3 m in altitude difference. By subtracting the standard QNH value at sea level, we can derive a rough MSL height and subsequently convert MSL to HAE.

\subsubsection{Summary}
The proposed framework, as illustrated in~\cref{fig: silas_core}, offers a practical framework for bridging discrepancies between height systems in lower airspace, facilitating compatibility and interoperability while maintaining an acceptable level of precision.
While this framework does not yet resolve all challenges inherent to height system unification, it provides a balanced compromise between competing demands. For example, transforming a legacy height (e.g., Barometric) to HAE via this framework facilitates interoperability but does not eliminate the inherent source error. The proposed framework aims to deliver a universal solution without introducing unnecessary complexity. For systems like barometric altitude, where inherent inaccuracies have persisted for decades, a definitive resolution remains elusive. However, a promising path forward may lie in crowd-sourced data collection or establishing a dense correction look-up table using numeric weather prediction data~\cite{schumann2017static,Simonetti2024GeodeticAF}. By aggregating large-scale, site-specific pairwise measurements (e.g., barometric vs. geometric heights), a dense calibration network could be established to enable data-driven refinements of empirical models.


\section{From Concept to Reality: Case Studies in Shenzhen}
\label{sec: airspace}

To validate the operational efficacy of the HAE framework, we present two complementary case studies that address the primary constraints of lower airspace: \textit{static terrain obstacles} and \textit{dynamic traffic density}.

\textbf{Case 1 (Static Safety)} focuses on the \textit{spatial definition} of airspace. Using high-resolution DEM data from Shenzhen, we demonstrate how HAE enables precise, data-secure zoning of navigable airspace relative to terrain, resolving the ambiguities of AGL.

\textbf{Case 2 (Dynamic Capacity)} focuses on the \textit{throughput utilization} of that airspace. By analyzing sensor error budgets, we quantify how HAE's reference stability allows for tighter vertical stacking of flight corridors, transforming airspace from a scarce resource into a scalable infrastructure.

\subsection{Case 1: Partitioned Airspace Management Using HAE}

\subsubsection{Current Status}
\begin{figure}
    \centering
    \includegraphics[width=1\linewidth]{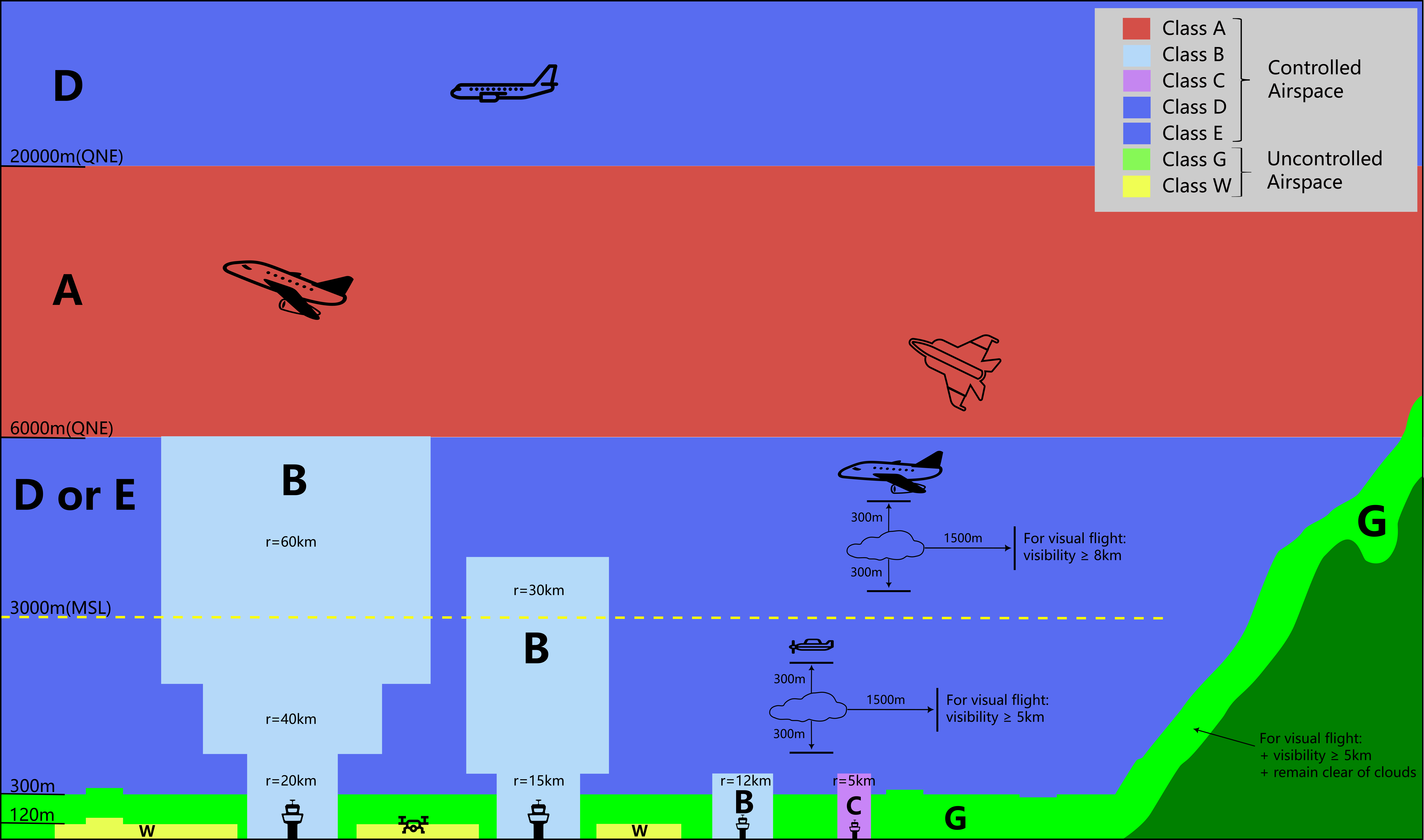}
    \caption{Sketch of current airspace management by authorities.}
    \label{fig:enter-airspace}
\end{figure}

Airspace management is a complex issue that remains unresolved, although various works have been proposed to analyze and address it~\cite{bauranov2021designing}. According to existing frameworks like the ``Regulations on Airspace Management''~\cite{caac2023yjs} and ``National Airspace Basic Classification Method''~\cite{caac2023}, airspace is divided into seven classes: A, B, C, D, E, G, and W, each based on different factors such as aircraft flight rules, performance requirements, the airspace environment, and air traffic control services. Each class corresponds to distinct height levels, types of airspace, and specific aircraft flight requirements, as depicted in~\cref{fig:enter-airspace}. Class G and class W represent uncontrolled airspace, while others fall under controlled airspace regulations. In low-altitude flight scenarios, the focus of lower airspace lies mainly on class G and class W:
\begin{itemize}
    \item Class G: Includes airspace outside of class B and class C, with a true height below 300 meters (excluding class W airspace), and airspace below 6000 meters above mean sea level that does not affect public civil transport flights.
    \item Class W: Includes portions of class G airspace with a true height below 120 meters.
\end{itemize}
Although the regulations attempt to classify airspace based on height systems, the descriptions provided remain insufficiently clear and even problematic in certain respects. For example, the document fails to explicitly define the meaning of ``true height'', and the accompanying figure legend seems incomplete. The left side of the figure omits the representation of buildings and terrain, yet the W and G airspace show slight uplifts and depressions. This suggests that the depicted height is based on DSM-derived AGL, while others argue that, according to international convention, true height should strictly refer to the DTM-derived AGL. However, the authorities have yet to offer clarification on this issue. This ambiguity regarding the AGL reference datum is not unique to Chinese regulations and similar issues arise in systems like LAANC. 
Furthermore, the restriction that class G airspace is limited to ``areas below 6000 meters mean sea level'' seems somewhat illogical. While regions with an MSL above 6000 meters are indeed rare, prohibiting UAV usage in high-altitude areas appears counterintuitive. These ambiguities in this document introduce uncertainty regarding the safe operation of UAVs and other aircraft in uncontrolled airspace, thereby hindering lawful flight operations and impeding LAE development.

\subsubsection{Proposed Partitioned Airspace Management Solution}
Setting aside the ambiguities in the document, let us assume that the height system in question does indeed adhere to the AGL. Under such an assumption, a critical feasibility issue arises: how can users reliably determine regulatory compliance? This hinges on two unresolved challenges: (1) the technical difficulty of obtaining precise AGL measurements, and (2) the inability of users to verify compliance due to restricted access to high-precision DTM data, as outlined in~\cref{sec:preliminaries}. While AGL theoretically reflects aircraft-to-ground proximity for safety, its reliance on confidential DTM data renders it impractical for real-world regulation. Without open access to DTMs, AGL-based systems remain conceptual rather than actionable. 

Despite the practical challenges of implementing AGL as a height management system, it does offer inherent advantages. Since human activities primarily take place at or near the Earth's surface, AGL directly reflects the distance between the surface and aircraft, making it a reasonable basis for ensuring low-altitude safety. 
To preserve AGL’s safety benefits while ensuring operational viability, we propose a partitioned airspace management framework using HAE. By partitioning regions into zones with uniform elevation thresholds (based on terrain homogeneity), HAE addresses AGL’s limitations while offering distinct advantages:
\begin{itemize} 
    \item \textbf{Clarity}: HAE derives from a mathematically defined global ellipsoid, bypassing ambiguous geoid-based vertical datums and terrain-dependent references. This standardization ensures singular interpretability--all stakeholders such as airspace users and regulators share identical definitions of airspace boundaries, eliminating discrepancies inherent to terrain-dependent references.
    \item \textbf{Security}: HAE requires no sensitive DTM data. Using HAE for height management eliminates the need for users to access DTM data, thereby addressing potential security risks. Moreover, GNSS technology is highly mature and widely utilized across civilian and military sectors, ensuring that the use of HAE in cross-border or large-area airspace management does not pose risks of information leakage or security breaches.
    \item \textbf{Interoperability}: HAE serves as a universal intermediary, easily convertible to other height systems for seamless integration with legacy systems. This allows airspace users using legacy systems to verify compliance with airspace management regulations.
    \item \textbf{Transparency}: GNSS accessibility ensures equitable usage, allowing any user to independently verify HAE, thus avoiding jurisdiction-specific data barriers.
\end{itemize}
The next critical issue, therefore, is how to effectively delineate management zones and establish definitive HAE height thresholds, which is detailed as follow.
\begin{figure*}[!htbp]
    \centering
    \includegraphics[width=1\linewidth]{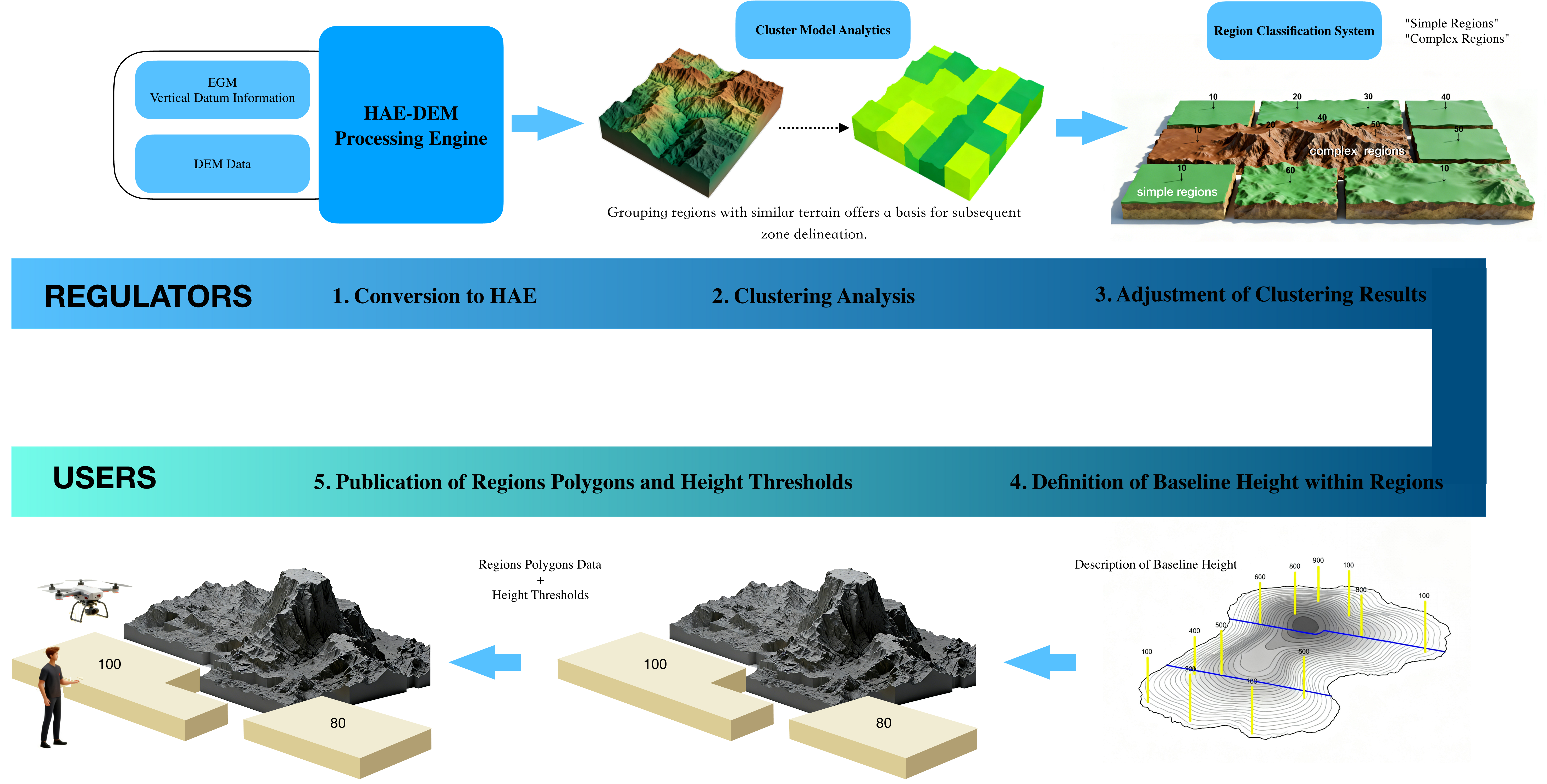}
    \caption{Implementation process of the airspace management scheme based on HAE.}
    \label{fig:management}
\end{figure*}
The partitioned airspace management framework necessitates minimizing terrain variation within zones. Clustering analysis offers the most effective solution: by grouping regions with similar elevation characteristics, statistical measures (e.g., variance, percentiles) from each cluster can establish baseline height thresholds for unregulated airspace. To align with HAE’s operational advantages (universal accessibility, GNSS compatibility) and ensure practical implementation, we propose the following systematic workflow in~\cref{fig:management}:
\begin{enumerate}

    \item Conversion to HAE: DEM data are typically represented in raster format, reflecting MSL. We can leverage known data of the vertical datum to get its relative to the Earth ellipsoid, or obtain geoid heights for each raster pixel point through EGM to transform the original DEM raster data into a HAE described raster data.

    \item Clustering Analysis: Given that terrain changes continuously across space, adjacent pixels exhibit similar height values. Thus, even without considering spatial proximity, employing simple clustering algorithms such as K-means can yield spatially continuous clustering results. This method effectively groups regions with similar terrain characteristics, providing a rational basis for subsequent zone delineation.

    \item Adjustment of Clustering Results: To make the published height thresholds more intuitive and convenient, the height values for each cluster can be rounded to whole tens. This adjustment facilitates public recall and application of height information while simplifying calculations and communication in management and monitoring processes. Furthermore, based on the terrain characteristics, clusters can be classified into ``simple'' and ``complex'' regions. Regions with relatively flat terrain will be designated as simple regions, while more rugged terrains, such as mountainous regions, will be classified as complex regions. The final result will consist of multiple simple areas and one complex area, reflecting the inherent differences in terrain across the region.

    \item Definition of Baseline Height within Regions: Within each region, a baseline height must be established to define the corresponding class W airspace and class G airspace. For simple regions, which generally cover larger areas, this baseline can be derived from statistical measures of all height values within the region, such as the median or the 75th percentile. In contrast, for complex regions—typically smaller and with more varied terrain—baseline heights can be specified at regular intervals , such as every 100 meters, rather than relying on a single statistical measure, providing more practicality and flexibility.

    \item Publication of Polygons and Height Thresholds: Finally, polygons representing the areas corresponding to the clustering results and the HAE height thresholds for each area can be published. This information will provide clear height guidance for the public, assisting users in adhering to regulations in uncontrolled areas and supporting the healthy development of lower airspace activities.

\end{enumerate}

\subsubsection{A preliminary experiment \& analysis}

\begin{figure}[!htbp]
    \centering
    \includegraphics[width=1\linewidth]{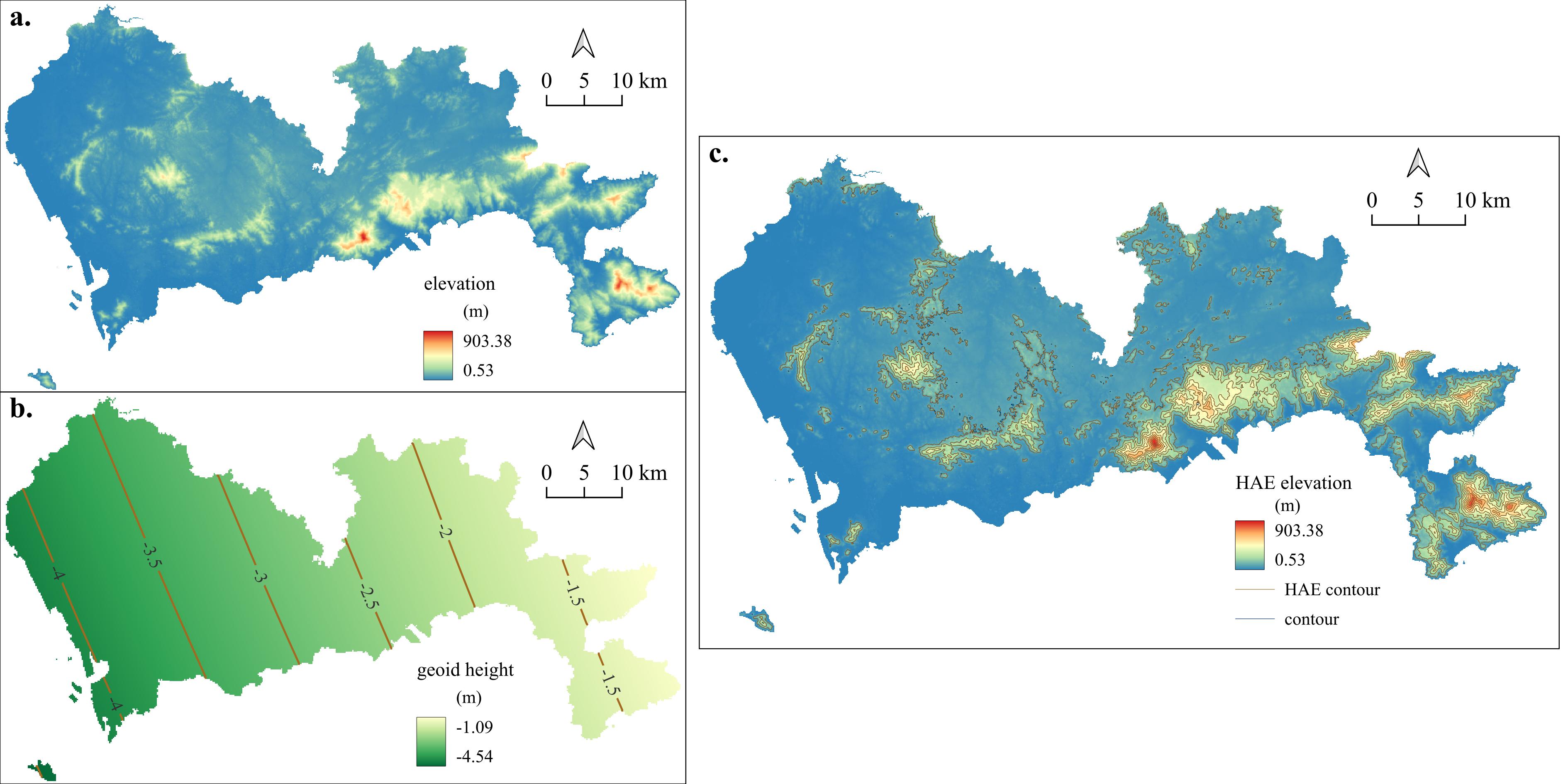}
    \caption{Overview of elevation, geoid height, and HAE in Shenzhen.}
    \label{fig:hae-app1}
\end{figure}

To demonstrate the operability of the aforementioned process, we applied it to the MERIT dataset focusing on the Shenzhen area. As previously mentioned, the MERIT dataset, which has been cleared of features such as trees and buildings, is more representative of a DTM, corresponding to the EGM96. The clipped elevation data from the MERIT dataset, the corresponding geoid heights for the region, and the computed HAE elevations are illustrated in the figures. In~\cref{fig:hae-app1}, contour lines at 100-meter intervals were generated based on both the original elevation data and the HAE elevations. Although the geoid heights are relatively low, there are still slight differences in the contours between the two datasets.

\begin{figure}[!htbp]
    \centering
    \includegraphics[width=1\linewidth]{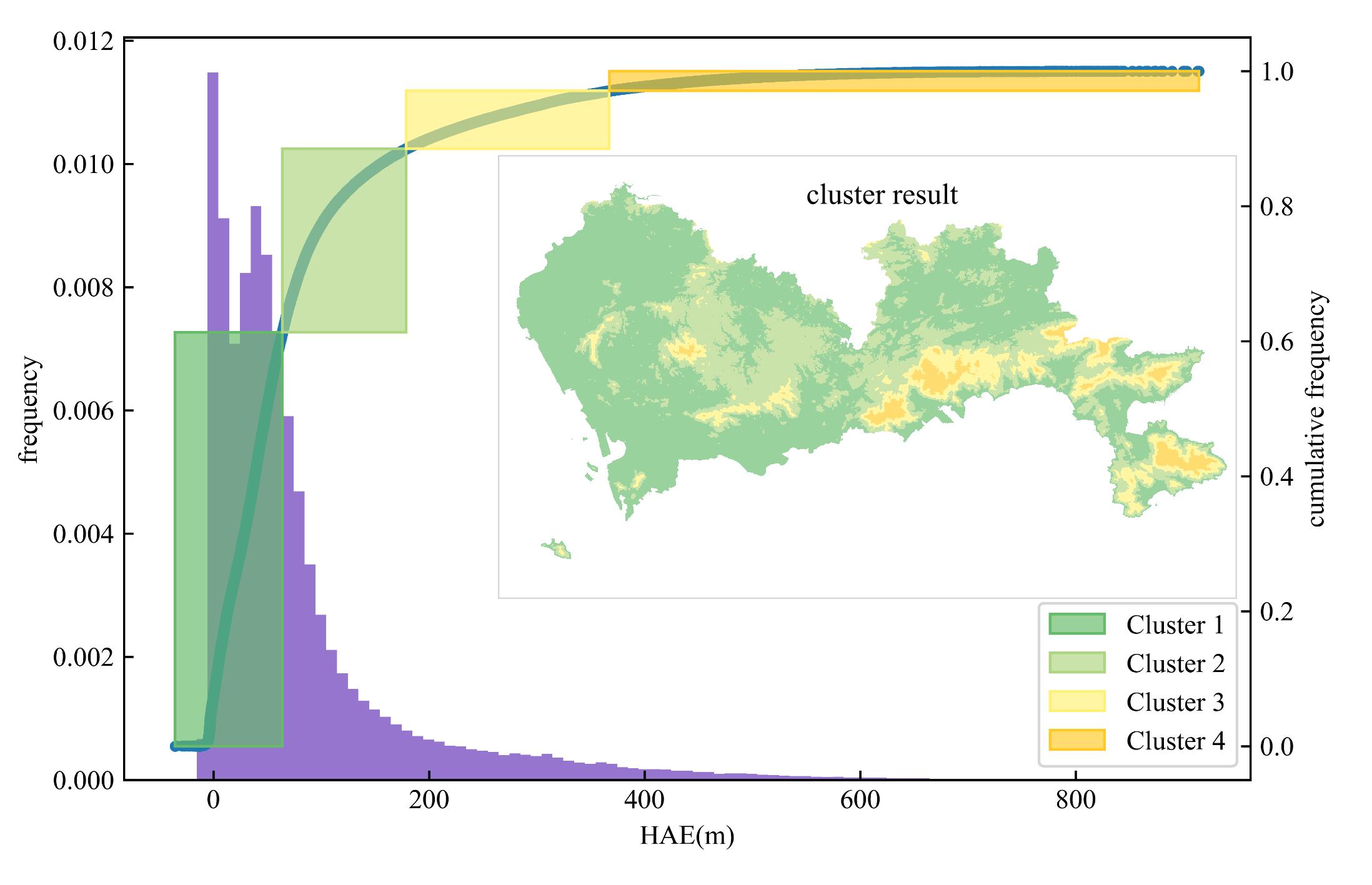}
    \caption{Clustering results of HAE.}
    \label{fig:hae-app2}
\end{figure}

Overall, K-means clustering with k=4 was performed on the HAE values for Shenzhen, where k=4 was determined using the Elbow Method. The results are as shown in~\cref{fig:hae-app2}. The histogram of HAE values indicates that they roughly follow a power law distribution, with the values concentrated in the lower range. The maximum HAE values for the four clusters are 63.96 m, 178.90 m, 367.19 m, and 914.26 m, respectively. Clusters 1 and 2 correspond to lower HAE ranges, collectively covering nearly 90\% of the area (with Cluster 1 comprising 61.28\% and Cluster 2 27.2\%). Clusters 3 and 4 correspond to higher value ranges, accounting for only 8.57\% and 2.95\% of the area, respectively. Based on the clustering results, Clusters 1 and 2 can be designated as simple regions, while Clusters 3 and 4 can be aggregated into a complex region.

\begin{figure}[!htbp]
    \centering
    \includegraphics[width=1.0\linewidth]{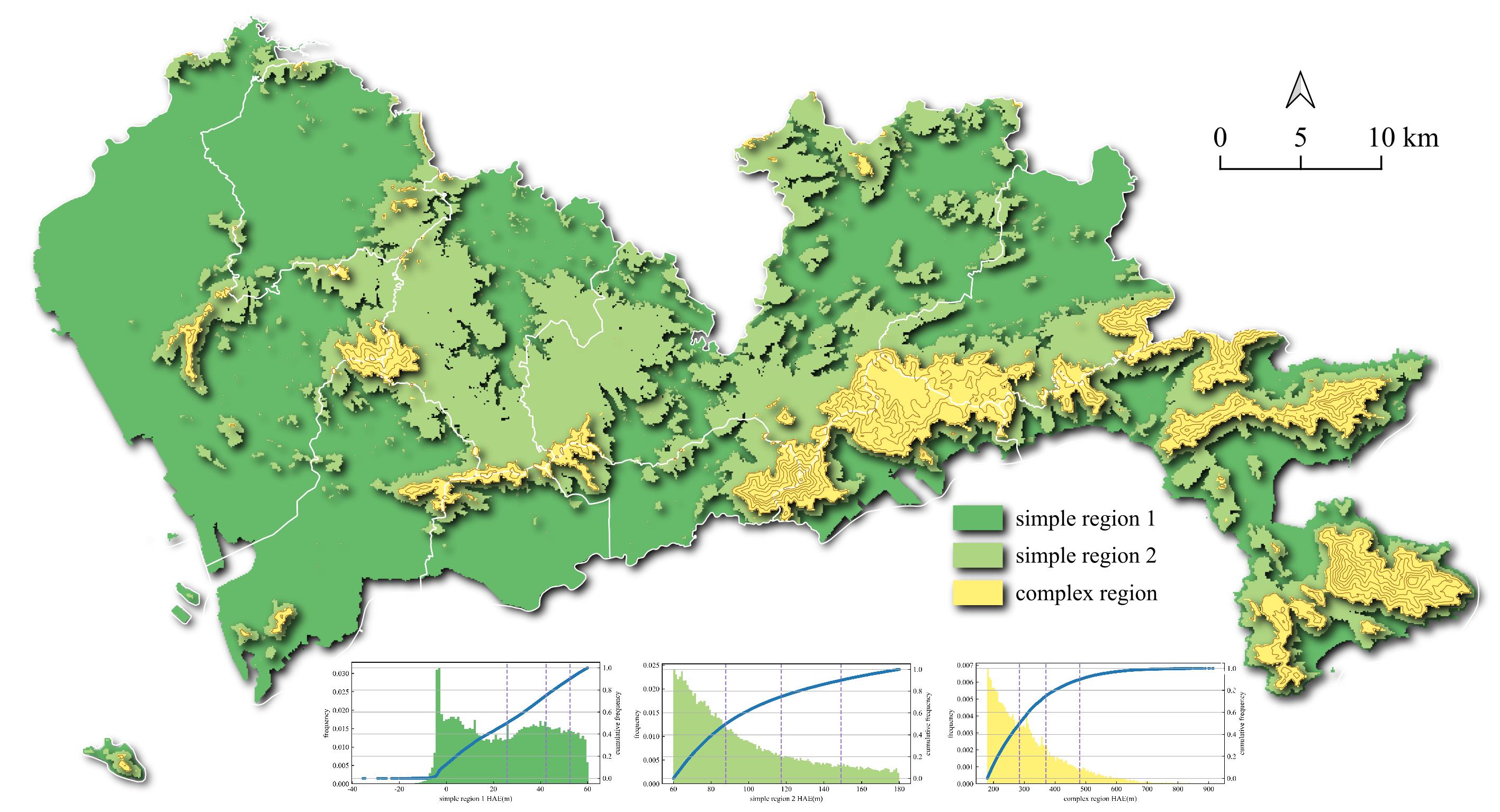}
    \caption{Classification of simple and complex regions.}
    \label{fig:hae-app3}
\end{figure}
\begin{table*}[!htbp]
    \centering
    \caption{Analysis of Height Data by Region}
    \begin{tabular}{|l|r|r|r|r|r|r|r|r|}
        \hline
        \small
        \textbf{Region Category} & \textbf{Pixel Count} & \textbf{Mean (m)} & \textbf{Std} & \textbf{Q1 (m)} & \textbf{Median (m)} & \textbf{Q3 (m)} & \textbf{Skew} & \textbf{Kurtosis} \\
        \hline
        Simple Region 1 & 153699 & 25.31 & 19.71 & 6.96 & 25.79 & 42.37 & 0.04 & -1.28 \\
        \hline
        Simple Region 2 & 78755 & 97.56 & 31.64 & 71.61 & 87.85 & 117.28 & 0.87 & -0.28 \\
        \hline
        Complex Region & 30015 & 314.65 & 117.48 & 223.94 & 284.95 & 370.85 & 1.34 & 1.94 \\
        \hline
    \end{tabular}%
    \label{tab:height_data_analysis}
\end{table*}

The rounded thresholds of 60 m and 180 m are established. Specifically, Simple Region 1 corresponds to areas with HAE not exceeding 60 m, Simple Region 2 corresponds to areas with HAE between 60 m and 180 m, and the complex region encompasses areas with HAE greater than 180 m. The aggregated regions and the internal distribution of HAE are illustrated as shown in~\cref{fig:hae-app3} and related statistics are shown in~\cref{tab:height_data_analysis}.

Clearly, there are differences in data distribution between the simple and complex regions. Both simple regions exhibit kurtosis values less than 0, while the complex region has a kurtosis greater than 0. This suggests that the data distribution in the simple regions is flatter and more spread out, while the complex region shows a more peaked distribution with heavier tails. The skewness of Simple Region 1 approaches 0, indicating that its data distribution is relatively symmetric, suggesting that the HAE values in this area vary within a narrow range without significant skewness. Based on the statistical analysis, it is apparent that the clustering method has effectively captured the terrain variability and provided a clear division of the study area into simple and complex regions, each with distinct height characteristics.

The next step involves determining a baseline height for Simple Region 1 and Simple Region 2, as well as establishing an appropriate interval for the Complex Region. This process can be somewhat subjective and may necessitate the inclusion of additional data; for example, population density or aircraft activity data could help refine the selection of baseline heights and intervals.

For simplicity, the median height is employed as the baseline for the simple regions, rounded to 25m and 90m, respectively. This choice reflects the relatively uniform terrain and stable HAE characteristics of the simple regions, making the median a suitable representative value. In contrast, a 100m interval is applied to the complex region, where terrain variability is more pronounced. Here, using the maximum value of each interval as the baseline height is appropriate because it accounts for the more complex and varied topography, ensuring that the airspace meets the requirements for flight safety.

Consequently, the maximum HAE limits for class W airspace in Shenzhen's simple regions are 145m and 210m. For the complex region, divisions are made at 100m intervals, with the maximum value of each interval serving as the baseline height. Based on this methodology, the corresponding maximum HAE for each segmented complex region of Class W airspace is determined. For instance, if 300m HAE is the maximum height in a segment, then the corresponding maximum HAE for the Class W airspace in that region would be 420m.

Finally, the management authority can release the polygons corresponding to each region and the HAE restrictions for the uncontrolled airspace based on the results of the aforementioned process, without involving any sensitive data operations. Users can determine the type of area in which the aircraft is located through relevant services and verify whether the aircraft is in compliance by obtaining its HAE via GNSS, without relying on any external data. Once this framework is established, the operations of various low-altitude aircraft can proceed more smoothly, thereby laying a solid foundation for the realization of LAE. Furthermore, the HAE-based framework can be converted to any other height system through the aforementioned conversion rules, enabling the release of the same data in different height systems without the need for additional equipment or operations, depending on the specific requirements of various stakeholders.

\subsection{Case 2: Safety and Capacity Analysis via Empirical Error Modeling}
\label{sec:case2}
To rigorously evaluate the operational performance of the proposed HAE framework against legacy pressure-based systems, we moved beyond theoretical assumptions and conducted a probabilistic risk assessment driven by empirical data. This study quantifies the impact of vertical reference stability on two critical metrics: safety (Vertical Separation Minimum) and efficiency (Airspace Throughput), referencing standards from\cite{docmanual}.

\subsubsection{Empirical Uncertainty Extraction}
We utilized real-world flight logs from the PX4 Autopilot ecosystem~\footnote{\url{https://review.px4.io/}} to characterize the error distributions of both height systems. The dataset consists of flight logs from multicopters equipped with RTK-GNSS modules, operating in diverse low-altitude environments.

\textbf{Ground Truth Generation:} The ellipsoidal height ($h_{RTK}$) derived from the Real-Time Kinematic (RTK) solution was treated as the ground truth for vertical positioning, owing to its centimeter-level precision.

\begin{figure}[!t]
\centering
\includegraphics[width=\linewidth]{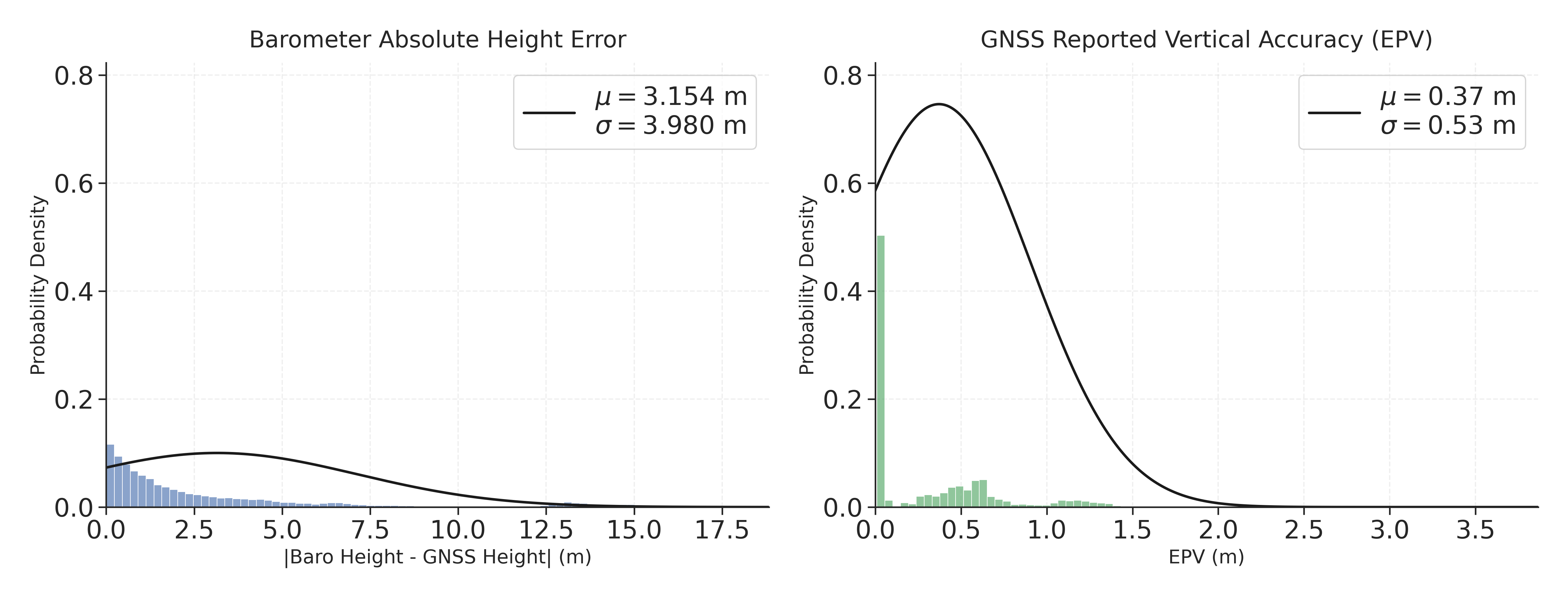} 
\caption{Empirical Error Distribution derived from PX4 Flight Logs. The legacy barometer (Left) exhibits significant variance ($\sigma \approx 3.98$~m) relative to RTK ground truth. The GNSS HAE model (Right), based on reported EPV, maintains high precision ($\sigma \approx 0.53$~m).}
\label{fig:error_model}
\end{figure}

\textbf{Barometric Error Modeling:} We extracted the raw barometric altitude ($h_{baro}$) and compared it against the RTK ground truth. For each flight segment, the initial bias was removed to simulate a standard pre-flight QNH calibration. The residual error, defined as $\varepsilon_{baro} = h_{baro} - h_{RTK}$, represents the in-flight drift and sensor noise. As shown in Fig. \ref{fig:error_model} (Left), the barometric error exhibits a wide dispersion with a standard deviation of $\sigma_{baro} \approx 3.98$~m, attributed to atmospheric micro-variations and sensor drift.

\textbf{GNSS HAE Error Modeling:} To model the uncertainty of the proposed HAE system, we excluded the RTK measurements used for ground truth. Instead, we analyzed the Estimated Position Error Vertical (EPV) recorded in the PX4 logs. The EPV represents the system's internal confidence in its vertical position estimate, serving as a realistic proxy for the performance of standard consumer-grade GNSS under good satellite visibility. As shown in Fig. \ref{fig:error_model} (Right), the HAE uncertainty is tightly constrained, with a standard deviation of $\sigma_{HAE} \approx 0.53$~m.

\subsubsection{Vertical Collision Risk Analysis (Reich Model)}
\begin{figure}[!t]
\centering
\includegraphics[width=\linewidth]{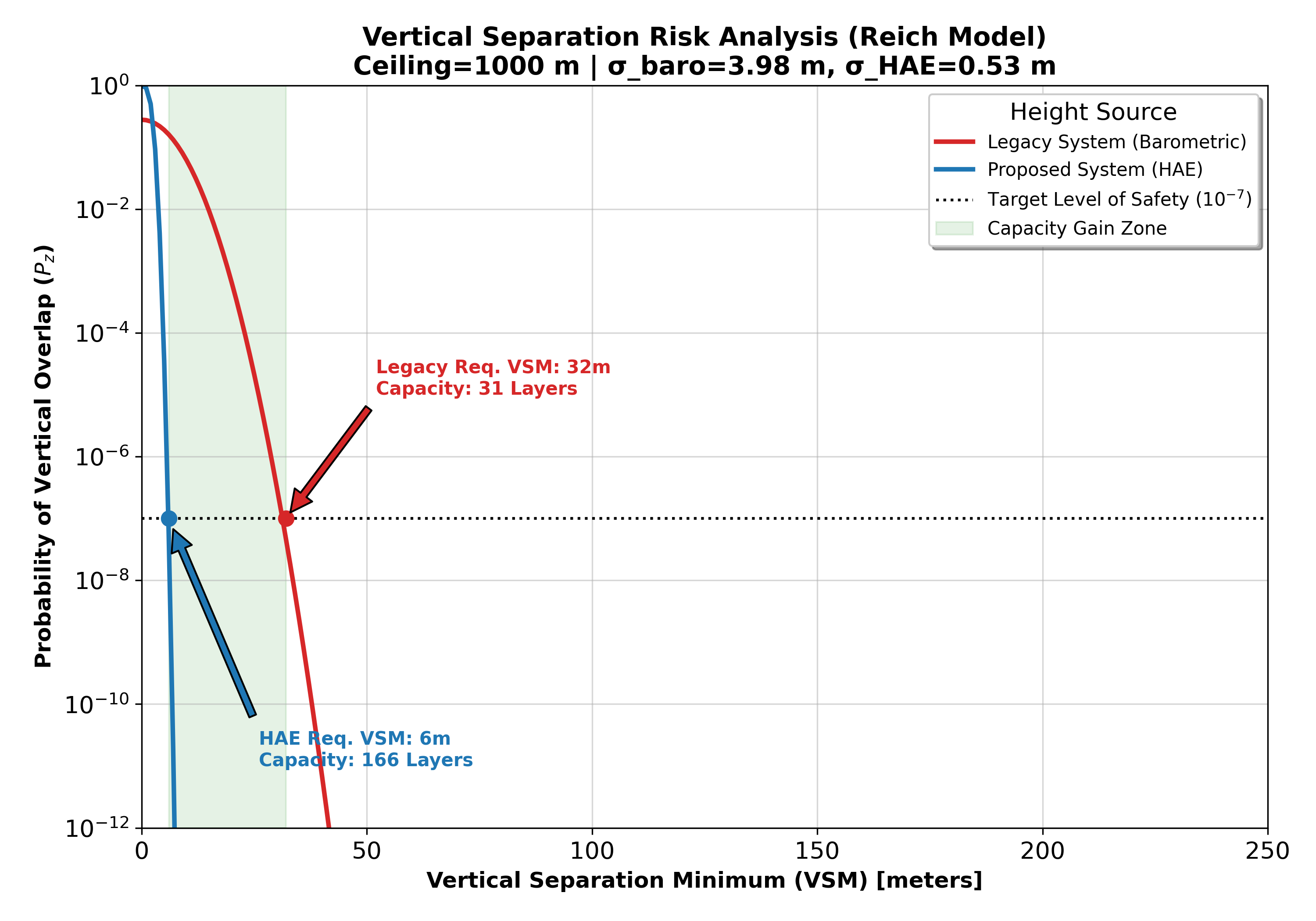} 
\caption{Vertical Separation Risk Analysis using the Reich Collision Risk Model \cite{Reich1966}. The required VSM is computationally derived where the overlap probability intersects the TLS ($10^{-7}$).}
\label{fig:reich_model}
\end{figure}

Based on the extracted empirical error parameters ($\sigma_{baro}$ and $\sigma_{HAE}$), we applied the Reich Collision Risk Model (CRM) to determine the necessary safe separation distances. The Probability of Vertical Overlap ($P_z$) between two aircraft assigned to adjacent flight levels separated by distance $S$ is calculated via:

\begin{equation}
P_z(S) = \int_{-\infty}^{\infty} f_{err}(z) \cdot f_{err}(z - S) \, dz
\end{equation}

where $f_{err}$ represents the probability density function of the vertical error derived from the empirical data.  Rather than selecting arbitrary separation distances, we computed the required VSM using the collision risk methodology defined in~\cite{docmanual}. The VSM is derived to ensure that the probability of vertical overlap does not exceed a Target Level of Safety (TLS) of $10^{-7}$.
The required separation $S$ is computed as:
\begin{equation}
S = \lambda \cdot \sqrt{\sigma_{1}^2 + \sigma_{2}^2} = \lambda \cdot \sqrt{2}\sigma
\label{eq:rvsm}
\end{equation}
where $\sigma$ is the empirical standard deviation derived in the previous section, and $\lambda$ is the safety factor corresponding to the TLS. For a TLS of $10^{-7}$, $\lambda \approx 5.3$. Therefore, to maintain the TLS, the legacy barometric system necessitates a Vertical Separation Minimum (VSM) of approximately {32 meters}. In contrast, the high precision of the HAE system allows the VSM to be safely compressed to {6 meters}. The results, illustrated in Fig. \ref{fig:reich_model}, demonstrate a dramatic disparity in safety requirements. In a standard 1,000-meter low-altitude airspace ceiling, this reduction in safety buffer translates directly into static capacity. The legacy system supports only {31 usable flight levels}, whereas the HAE framework unlocks {166 usable flight levels}, representing a $5.3\times$ increase in static airspace density.

\subsubsection{Operational Throughput Analysis (Erlang-B)}
To quantify the impact on dynamic logistics operations, we applied the Erlang-B loss model~\cite{erlang1948solution}, a standard metric in telecommunications and traffic engineering for blocking probability $P_b$:
\begin{equation}
P_b = \frac{\frac{A^N}{N!}}{\sum_{i=0}^{N} \frac{A^i}{i!}}
\end{equation}
where $N$ is the number of available flight levels (calculated as $\lfloor \text{Ceiling}/\text{VSM} \rfloor$) and $A$ is the traffic intensity (Erlangs). Here, the airspace was modeled as a resource system where usable flight levels function as ``servers'' and flight missions function as ``requests.'' We defined a Quality of Service (QoS) threshold where the probability of being denied service (due to airspace congestion) must not exceed 5\%.
\begin{figure}[!t]
\centering
\includegraphics[width=\linewidth]{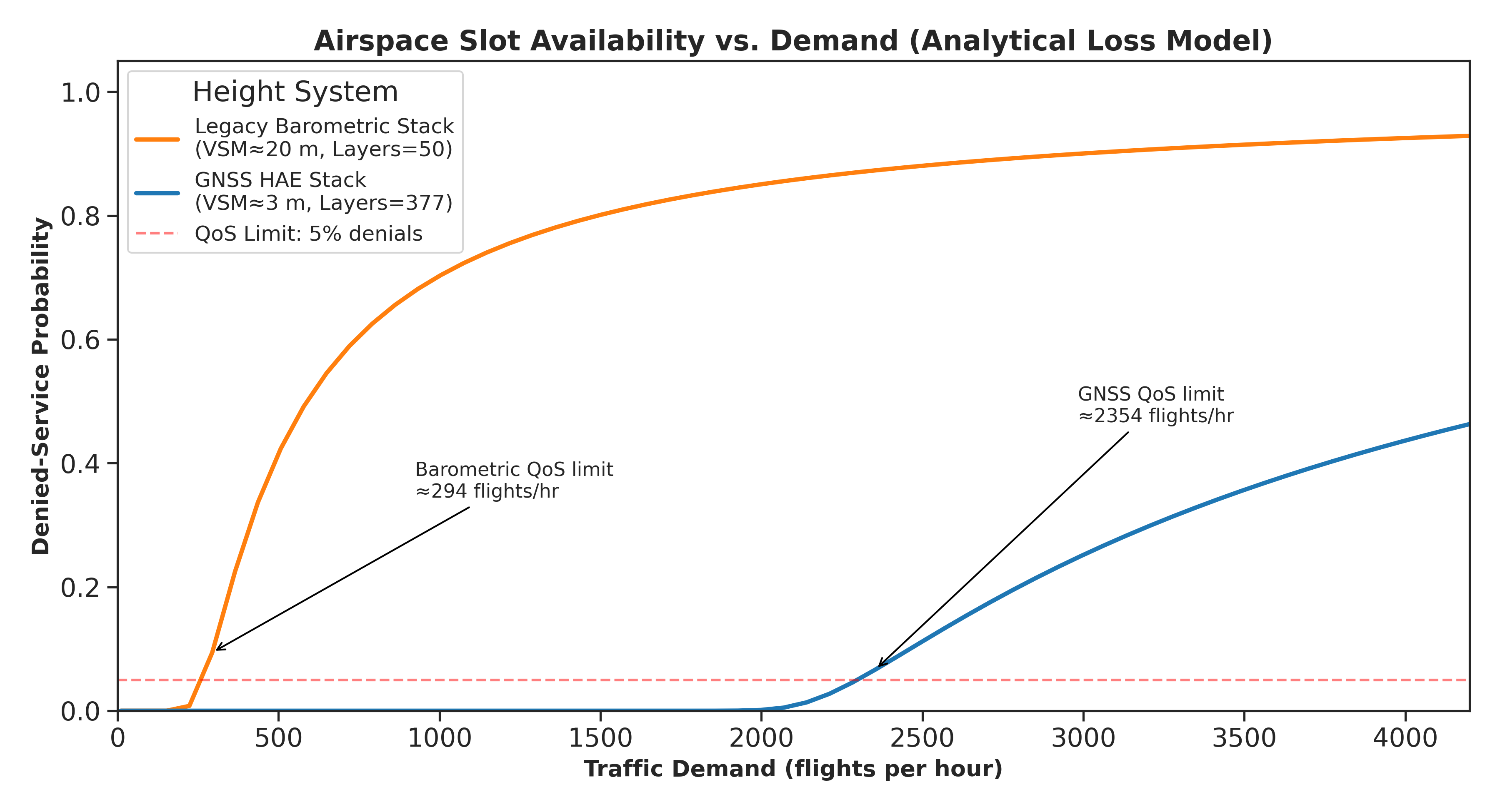} 
\caption{Airspace Slot Availability vs. Demand (Erlang-B Model). The proposed HAE system supports an $8\times$ increase in traffic throughput (2354 vs 294 flights/hr) before violating the 5\% Quality of Service (QoS) limit.}
\label{fig:capacity_gap}
\end{figure}

As presented in Fig. \ref{fig:capacity_gap}, the legacy barometric system saturates rapidly, reaching the QoS limit at a demand of approximately {294 flights per hour}. Conversely, the HAE system, leveraging its 166 available layers, sustains traffic demands up to {2,354 flights per hour} before reaching the same congestion threshold. This analysis confirms that transitioning to HAE provides an approximately {8-fold increase} in dynamic service capacity, validating it as a critical enabler for scalable low altitude activities.

\section{Conclusion and Discussion}
\label{sec:con}
In light of ongoing urban development and technological advancements, lower airspace management is attracting increasing scholarly and practical attention. Despite extensive research in air traffic management, the description of height—fundamental to airspace digitization—remains fragmented and ill-suited for high-density autonomous operations. This paper identifies Height Above Ellipsoid (HAE) as the optimal standardized vertical reference, offering superior digital compatibility, reference stability, and operational accessibility compared to legacy pressure-based or terrain-dependent systems. We proposed a bidirectional conversion framework to bridge legacy systems with HAE, providing a pragmatic pathway for modernization without discarding existing infrastructure. Furthermore, we introduced a partitioned airspace management framework anchored in HAE, which resolves the ambiguities of AGL by establishing standardized, accessible height thresholds for urban zoning. Crucially, our empirical validation employing the Reich Collision Risk Model demonstrates that the instability of legacy barometric systems necessitates prohibitive safety buffers ($\text{VSM} \approx 32$~m), severely constraining airspace utilization to just 31 usable flight levels within a 1,000~m ceiling. By stabilizing the vertical reference with GNSS-derived HAE, the safe separation buffer is compressed to 6~m, unlocking 166 usable flight levels. Dynamic capacity modeling further confirms that this precision translates to an 8-fold increase in traffic throughput (supporting over 2,300 flights/hour compared to just 294 with legacy systems) before reaching congestion limits.

In conclusion, the standardization of HAE is not merely a technical refinement but an economic necessity. It transforms lower airspace from a scarce, uncertain resource into a scalable digital infrastructure capable of supporting the trillion-dollar low-altitude economy. 
However, its widespread adoption faces two pivotal challenges:
\begin{itemize}
    \item Stakeholder Coordination: Diverse stakeholders including regulators, operators and manufacturers must align under unified HAE frameworks. This demands robust policy harmonization, phased certification processes, and iterative stakeholder engagement to balance legacy systems with modernization.
    \item Thorough Validation in Lower Airspace: While HAE leverages GNSS accuracy, dynamic low-altitude environments, e.g., urban canyons, signal multipath, require targeted research into real-time correction algorithms and redundancy mechanisms to ensure centimeter-level reliability.
\end{itemize}

As for future work, our future research will focus on:
\begin{itemize}
\item Implementation and Validation: Develop hardware prototypes and robust conversion algorithms to operationalize the HAE-legacy height system conversion scheme. This includes real-world testing to validate accuracy under dynamic low-altitude conditions.
    \item Multi-Criteria Partition Airspace Management: we plan to further enhance our partitioned airspace management solution by integrating safety, efficiency, and socio-environmental factors such as noise-sensitive zones and population density. Hybrid methods combining multi-objective optimization, machine learning, and geospatial clustering will be explored to balance zoning practicality and complexity.  
    \item Advance Geospatial Analysis: Collaborate with GIS communities to advance geospatial analytics using standardized crowdsourced flight data. Priorities include developing open 3D spatial query standards, 3D geospatial analysis, and creating adaptive digital twins for urban airspace management.  
\end{itemize}




\bibliographystyle{IEEEtran}
\bibliography{ref}


 





\end{document}